\documentclass[aps,prd,twocolumn,preprintnumbers,superscriptaddress,nofootinbib,floatfix]{revtex4-2}
\usepackage{amsmath}
\usepackage{amsfonts}
\usepackage{amssymb}

\usepackage{multirow}

\usepackage{graphicx}
\usepackage{color}

\usepackage[colorlinks=true,linkcolor=blue,citecolor=blue,urlcolor=blue]{hyperref}

\usepackage{comment}

\newcommand{\result}[1]{#1}

\begin{document}


\newcommand{\oThreeRealHasMassGap}{\ensuremath{2.76\% \pm 0.09\%}}
\newcommand{\oThreeRealHasNS}{\ensuremath{2.25\% \pm 0.06\%}}
\newcommand{\oThreeRealBNS}{\ensuremath{1.21\% \pm 0.03\%}}
\newcommand{\oThreeRealNSBH}{\ensuremath{1.03\% \pm 0.05\%}}
\newcommand{\oThreeRealBBH}{\ensuremath{97.75\% \pm 0.25\%}}
\newcommand{\oThreeRealBBHa}{\ensuremath{2.80\% \pm 0.09\%}}
\newcommand{\oThreeRealBBHb}{\ensuremath{5.92\% \pm 0.16\%}}
\newcommand{\oThreeRealBBHc}{\ensuremath{19.53\% \pm 0.39\%}}
\newcommand{\oThreeRealBBHd}{\ensuremath{29.78\% \pm 0.54\%}}
\newcommand{\oThreeRealBBHe}{\ensuremath{39.72\% \pm 0.55\%}}
\newcommand{\oThreeRealqa}{\ensuremath{0.46\% \pm 0.07\%}}
\newcommand{\oThreeRealqb}{\ensuremath{8.97\% \pm 0.31\%}}
\newcommand{\oThreeRealqc}{\ensuremath{29.99\% \pm 0.50\%}}
\newcommand{\oThreeRealqd}{\ensuremath{60.58\% \pm 0.55\%}}
\newcommand{\oThreeSemiHasMassGap}{\ensuremath{2.67\% \pm 0.04\%}}
\newcommand{\oThreeSemiHasNS}{\ensuremath{2.55\% \pm 0.04\%}}
\newcommand{\oThreeSemiBNS}{\ensuremath{1.39\% \pm 0.03\%}}
\newcommand{\oThreeSemiNSBH}{\ensuremath{1.16\% \pm 0.03\%}}
\newcommand{\oThreeSemiBBH}{\ensuremath{97.45\% \pm 0.22\%}}
\newcommand{\oThreeSemiBBHa}{\ensuremath{2.60\% \pm 0.04\%}}
\newcommand{\oThreeSemiBBHb}{\ensuremath{5.42\% \pm 0.06\%}}
\newcommand{\oThreeSemiBBHc}{\ensuremath{17.13\% \pm 0.10\%}}
\newcommand{\oThreeSemiBBHd}{\ensuremath{28.56\% \pm 0.13\%}}
\newcommand{\oThreeSemiBBHe}{\ensuremath{43.73\% \pm 0.16\%}}
\newcommand{\oThreeSemiqa}{\ensuremath{0.69\% \pm 0.02\%}}
\newcommand{\oThreeSemiqb}{\ensuremath{9.01\% \pm 0.08\%}}
\newcommand{\oThreeSemiqc}{\ensuremath{29.96\% \pm 0.13\%}}
\newcommand{\oThreeSemiqd}{\ensuremath{60.34\% \pm 0.18\%}}
\newcommand{\oFourLowHasMassGap}{\ensuremath{3.16\% \pm 0.04\%}}
\newcommand{\oFourLowHasNS}{\ensuremath{2.97\% \pm 0.04\%}}
\newcommand{\oFourLowBNS}{\ensuremath{1.60\% \pm 0.03\%}}
\newcommand{\oFourLowNSBH}{\ensuremath{1.37\% \pm 0.03\%}}
\newcommand{\oFourLowBBH}{\ensuremath{97.03\% \pm 0.21\%}}
\newcommand{\oFourLowBBHa}{\ensuremath{2.98\% \pm 0.04\%}}
\newcommand{\oFourLowBBHb}{\ensuremath{6.17\% \pm 0.06\%}}
\newcommand{\oFourLowBBHc}{\ensuremath{18.23\% \pm 0.10\%}}
\newcommand{\oFourLowBBHd}{\ensuremath{28.26\% \pm 0.13\%}}
\newcommand{\oFourLowBBHe}{\ensuremath{41.37\% \pm 0.15\%}}
\newcommand{\oFourLowqa}{\ensuremath{0.74\% \pm 0.02\%}}
\newcommand{\oFourLowqb}{\ensuremath{9.07\% \pm 0.07\%}}
\newcommand{\oFourLowqc}{\ensuremath{30.18\% \pm 0.13\%}}
\newcommand{\oFourLowqd}{\ensuremath{60.02\% \pm 0.18\%}}
\newcommand{\oFourHighHasMassGap}{\ensuremath{3.32\% \pm 0.04\%}}
\newcommand{\oFourHighHasNS}{\ensuremath{3.14\% \pm 0.04\%}}
\newcommand{\oFourHighBNS}{\ensuremath{1.71\% \pm 0.03\%}}
\newcommand{\oFourHighNSBH}{\ensuremath{1.43\% \pm 0.03\%}}
\newcommand{\oFourHighBBH}{\ensuremath{96.86\% \pm 0.21\%}}
\newcommand{\oFourHighBBHa}{\ensuremath{3.15\% \pm 0.04\%}}
\newcommand{\oFourHighBBHb}{\ensuremath{6.44\% \pm 0.06\%}}
\newcommand{\oFourHighBBHc}{\ensuremath{18.65\% \pm 0.10\%}}
\newcommand{\oFourHighBBHd}{\ensuremath{27.45\% \pm 0.12\%}}
\newcommand{\oFourHighBBHe}{\ensuremath{41.17\% \pm 0.15\%}}
\newcommand{\oFourHighqa}{\ensuremath{0.83\% \pm 0.02\%}}
\newcommand{\oFourHighqb}{\ensuremath{9.39\% \pm 0.07\%}}
\newcommand{\oFourHighqc}{\ensuremath{30.04\% \pm 0.13\%}}
\newcommand{\oFourHighqd}{\ensuremath{59.75\% \pm 0.17\%}}


\title{
Semianalytic Sensitivity Estimates for Catalogs of Gravitational-Wave Transients
}

\author{Reed Essick}
\email{essick@cita.utoronto.ca}
\affiliation{Canadian Institute for Theoretical Astrophysics, 60 St. George St, Toronto, Ontario M5S 3H8}
\affiliation{Department of Physics, University of Toronto, 60 St. George Street, Toronto, ON M5S 1A7}
\affiliation{David A. Dunlap Department of Astronomy, University of Toronto, 50 St. George Street, Toronto, ON M5S 3H4}

\begin{abstract}
    I investigate the sensitivity of gravitational-wave searches by analyzing the response of matched filters in stationary Gaussian noise.
    In particular, I focus on the ability to analytically model the distribution of observed filter responses maximized over coalescence phase and/or a template bank as well as the response of statistics defined for a network of detectors.
    Semianalytic sensitivity estimates derived assuming stationary Gaussian noise are compared to sensitivity estimates obtained from real searches processing real noise, which is neither perfectly stationary nor perfectly Gaussian.
    I find that semianalytic estimates are able to reproduce real search sensitivity for the LIGO-Virgo-KAGRA Collaboration's third observing run with high fidelity.
    I also discuss how to select computational speed-ups (hopeless signal-to-noise ratio cuts) and make predictions for the fourth observing run using projected detector sensitivities.
\end{abstract}

\maketitle




\section{Introduction}
\label{sec:introduction}

Ensembles of detected Gravitational-Wave (GW) transients~\cite{GWTC-1, GWTC-2, GWTC-2.1, GWTC-3} encode the properties of the underlying astrophysical distribution of merging binaries~\cite{GWTC-2-RnP, GWTC-3-RnP} and the sensitivity of searches used to construct these catalogs.
Recently, the LIGO-Virgo-KAGRA (LVK) Collaborations released a set of injections (simulated signals) performed with real detector noise from the advanced LIGO~\cite{LIGO} and Virgo~\cite{Virgo} interferometers and real searches during their third observing run (O3~\cite{GWTC-3-injections}).
Earlier injections are also publicly available, but they span a smaller range of source-frame masses~\cite{O3a-injections}.
These injections are the gold-standard for sensitivity estimates, as they come closest to simulating the actual conditions in which searches detect GW transients and construct catalogs.
They can therefore be used to estimate search sensitivity to different types of coalescing binaries.
Various effects are relevant, but the strongest selection introduces a correlation in the detected distribution between a binary's (detector-frame) masses and the luminosity distance to the source.
As a rule of thumb, more massive binaries can be detected at larger distances.
Indeed, a solid understanding of selection effects alone can convey much of the important information within a catalog (see, e.g.,~\citet{Farah:2020}).

However, injection campaigns are notoriously expensive, requiring significant person-power and computing time to prepare and process large numbers of injections.
The effort for O3 involved searches processing \result{$> 1.1$ million} injections (\result{$> 86$ million} injections were generated). 
See Sec.~\ref{sec:real injections} for more discussion of search results for this injection set.
As such, several authors have proposed alternatives.

Typically, these rely on approximations for properties of detector noise and the actual search's behavior.
One can then emulate which injections would have been detected with more computationally efficient calculations.
\citet{Finn:1993} propose an approximation based on the optimal signal-to-noise ratio ($\rho_\mathrm{opt}$, Eq.~\ref{eq:rho opt}) in a single interferometer (IFO) to account for different binary orientations.
The detected population is modeled with a threshold on the single-IFO $\rho_\mathrm{opt}$.
This implicitly assumes the detector noise is stationary and Gaussian and that the Power Spectral Density (PSD) is known perfectly.
\citet{Fishbach:2019} and~\citet{GWTC-1-RnP} build upon this by modeling the observed signal-to-noise ratio ($\rho_\mathrm{obs}$) as a standard Gaussian random variate centered on $\rho_\mathrm{opt}$.
Selection is then made based on $\rho_\mathrm{obs}$ rather than $\rho_\mathrm{opt}$.
However, as we discuss below, $\rho_\mathrm{obs}$ is often not Gaussian distributed.

\citet{Tiwari:2017},~\citet{Veske:2021}, and~\citet{Gerosa:2020} propose similar techniques but consider the optimal signal-to-noise ratio from a network of detectors ($\rho_{\mathrm{net},\mathrm{opt}}$, Eq.~\ref{eq:rho net opt}) rather than a single IFO.
\citet{Gerosa:2020} explicitly states that they were motivated by possible failures in the single-IFO approximation.
Furthermore, they also assume that the binary's true parameters (masses, spins, etc.) are perfectly known.
However, searches do not actually witness $\rho_\mathrm{net,opt}$, nor do they know the binary's true parameters.
As~\citet{Fishbach:2019} attempted to capture, detector noise can scatter $\rho_\mathrm{obs}$ to both higher and lower values, and searches typically maximize a filter response over a broad template bank.
Additionally, real noise is neither perfectly stationary nor perfectly Gaussian and, even if it were, the PSD is not known perfectly.

Other authors have attempted to bootstrap injection sets.
However, such approximations axiomatically rely on pre-existing injection campaigns.
For example,~\citet{Talbot:2020} construct a Gaussian mixture model over a set of transformed single-event parameters (masses, spins, etc.) to emulate the distribution of detected binaries.
While this may provide an accurate emulator for a given set of injections, it may not readily generalize to other observing runs (different PSDs) or different detector networks.
It may be helpful, then, to develop an accurate emulator that can be trivially generalized to arbitrary detector networks with arbitrary sensitivities without having to generate new injections.
I claim that semianalytic sensitivity estimates based on $\rho_\mathrm{obs}$ can do exactly that.

It is therefore of some interest to consider how well one can approximate search sensitivity with estimates of the signal-to-noise ratio. 
I investigate the behavior of matched filter responses in the presence of stationary Gaussian noise and quantify the impact of common approximations to reduce the computational cost of constructing such estimates.
In Sec.~\ref{sec:wiener filter derivation}, I derive several matched filter signal-to-noise ratios that assume stationary Gaussian noise with a known PSD.
Sec.~\ref{sec:filter response distributions} investigates the how these statistics are distributed in the presence of an arbitrary signal, including the full covariance between the complex filter responses for an entire template bank.
Sec.~\ref{sec:sine-gaussian} studies how well one can approximate the distribution of $\rho_\mathrm{obs}$ maximized over an entire template bank with just the distribution of the response of template that best matches the true signal.
Sec.~\ref{sec:real injections} compares several semianalytic estimates with different levels of sophistication to real injections from O3~\cite{GWTC-3-injections} and makes projections for the LVK's fourth observing run (O4).
I then consider the impact of common computational shortcuts (``hopeless'' signal-to-noise ratio cuts) in Sec.~\ref{sec:hopeless cuts}.
I conclude in Sec.~\ref{sec:conclusion}.


\section{Matched filters in stationary Gaussian noise}
\label{sec:wiener filter derivation}

Consider additive, zero-mean stationary Gaussian noise ($n$) such that the observed data ($d$) in the presence of a signal ($h$) is given by
\begin{equation}
    d = n + h
\end{equation}
where $n$ is a random vector (function of frequency) described by a multivariate Gaussian distribution with
\begin{align}
    \mathrm{E}[\mathcal{R}\{n(f)\}]
        & = \mathrm{E}[\mathcal{I}\{n(f)\}] \nonumber \\
        & = 0 \\
    \mathrm{E}[\mathcal{R}\{n(f)\} \mathcal{R}\{n(f^\prime)\}]
        & = \mathrm{E}[\mathcal{I}\{n(f)\} \mathcal{I}\{n(f^\prime)\}] \nonumber \\
        & = \frac{1}{4} S(f) \delta(f - f^\prime) \\
    \mathrm{E}[\mathcal{R}\{n(f)\} \mathcal{I}\{n(f^\prime)\}]
        & = 0
\end{align}
where $\mathcal{R}\{\cdot\}$ and $\mathcal{I}\{\cdot\}$ denote the real and imaginary parts of a number.
This is often summarized as
\begin{gather}
    \mathrm{E}[n] = 0 \\
    \mathrm{E}[n(f) n^\ast(f^\prime)] = \frac{1}{2} S(f) \delta(f-f^\prime)
\end{gather}
where $(\cdot)^\ast$ denotes complex conjugation.
$S(f)$ is called the one-sided PSD.
This implies a corresponding likelihood in the frequency domain
\begin{align}
    \ln p(n) \supset -\frac{1}{2}\left( 4\int\limits_0^\infty df\,\frac{\mathcal{R}\{n(f)\}^2 + \mathcal{I}\{n(f)\}^2}{S(f)} \right)
\end{align}
so that the likelihood of observing $d$ given a signal $h$ is
\begin{align}
    \ln p(d|h) \supset -\frac{1}{2}\left( 4\int\limits_0^\infty df\, \frac{|d-h|^2}{S} \right)
\end{align}

If the spectral shape of a signal is known but the overall amplitude is note, then the maximum-likelihood estimate for the signal amplitude is the signal-to-noise ratio
\begin{equation}\label{eq:snr R}
    \rho_\mathcal{R}(h) = \frac{1}{\mathcal{N}(h)} \left(4\int\limits_0^\infty df\,\frac{\mathcal{R}\{d^\ast h\}}{S} \right)
\end{equation}
which is implicitly a function of both the data ($d$) and the template ($h$), with
\begin{equation}
    \mathcal{N}(h) \equiv \left( 4\int\limits_0^\infty df\,\frac{|h|^2}{S} \right)^{1/2}
\end{equation}
The process of calculating $\rho_\mathcal{R}$ is referred to as a matched filter. 
It is common to choose the template normalization so that $\mathcal{N}(h) = 1$ and neglect this term in Eq.~\ref{eq:snr R}.
I do not make this assumption and retain factors of $\mathcal{N}$ for clarity.

If the signal model is of the form $h=h_0 e^{i\phi}$, then it is also possible to maximize the filter response with respect to $\phi$ analytically.
The associated statistic is
\begin{equation}\label{eq:snr phi}
    \rho_{\phi}(h) = \left( \rho_\mathcal{R}^2 + \rho_\mathcal{I}^2 \right)^{1/2}
\end{equation}
where
\begin{equation}
    \rho_\mathcal{I}(h) = \frac{1}{\mathcal{N}(h)} \left(4\int\limits_0^\infty df\, \frac{\mathcal{I}\{d^\ast h\}}{S}\right)
\end{equation}
Similarly, one can maximize the response over an entire template bank
\begin{gather}
    \rho_{h,\mathcal{R}} = \max\limits_h \{\rho_\mathrm{R}(h)\} \\
    \rho_{h, \phi} = \max\limits_h \{\rho_\phi(h)\}
\end{gather}
although this must be done numerically, in general.

So far, these filter responses have been defined for a single detector.
One can define the equivalent statistics for a network of detectors by summing them in quadrature.
That is
\begin{align}
    \rho_{\mathrm{net}, \mathcal{R}}(h)
        & = \left( \sum\limits_i^{N_\mathrm{IFO}} \rho_{i,\mathcal{R}}^2(h) \right)^{1/2} \label{eq:snr net R} \\
    \rho_{\mathrm{net}, \phi}(h)
        & = \left( \sum\limits_i^{N_\mathrm{IFO}} \rho_{i,\phi}^2(h) \right)^{1/2} \label{eq:snr net phi}
\end{align}
This is because the maximized likelihood scales as $e^{+\rho^2/2}$ for each detector, and the assumption of independent noise implies the likelihood of the network will scale as $e^{+\sum_i \rho_i^2/2}$.
Note, however, that Eq.~\ref{eq:snr net phi} implicitly maximizes over $\phi$ separately for each detector and sums the results in quadrature.
Both Eqs.~\ref{eq:snr net R} and~\ref{eq:snr net phi} maximize over the signal amplitude independently in each detector.\footnote{While this is often done in practice, a real signal should appear consistently in all detectors, and one may instead maximize the quadrature sum of responses from all detectors simultaneously with respect to both the amplitude and $\phi$. I do not investigate the resulting statistics due to their considerable complexity and the fact that they are essentially unused in the field.}

One may additionally maximize the network responses over a template bank
\begin{align}
    \rho_{h, \mathrm{net}, \mathcal{R}}
        & = \max\limits_h \left\{ \rho_{\mathrm{net},\mathcal{R}}(h) \right\} \label{eq:snr h net R} \\
    \rho_{h, \mathrm{net}, \phi}
        & = \max\limits_h \left\{ \rho_{\mathrm{net},\phi}(h) \right\} \label{eq:snr h net phi}
\end{align}
Many searches based on matched filters will return something like Eq.~\ref{eq:snr h net phi}.\footnote{Some searches additionally impose consistency requirements for $\phi$ as inferred from different detectors in an attempt to account for the fact that the likelihood should not be optimized separately in each detector. See~\ref{sec:real injections} for more details about searches' behavior.}

This note examines the expected distributions of these statistics in the presence of stationary Gaussian noise.
In particular, I examine the ability to model these distributions analytically in Sec.~\ref{sec:filter response distributions} and~\ref{sec:sine-gaussian}.
I investigate how well those models reproduce the behavior of real searches in Sec.~\ref{sec:real injections}.


\begin{figure*}
    \begin{center}
        \Large{Single Detector Responses} \\
        \includegraphics[width=0.8\textwidth]{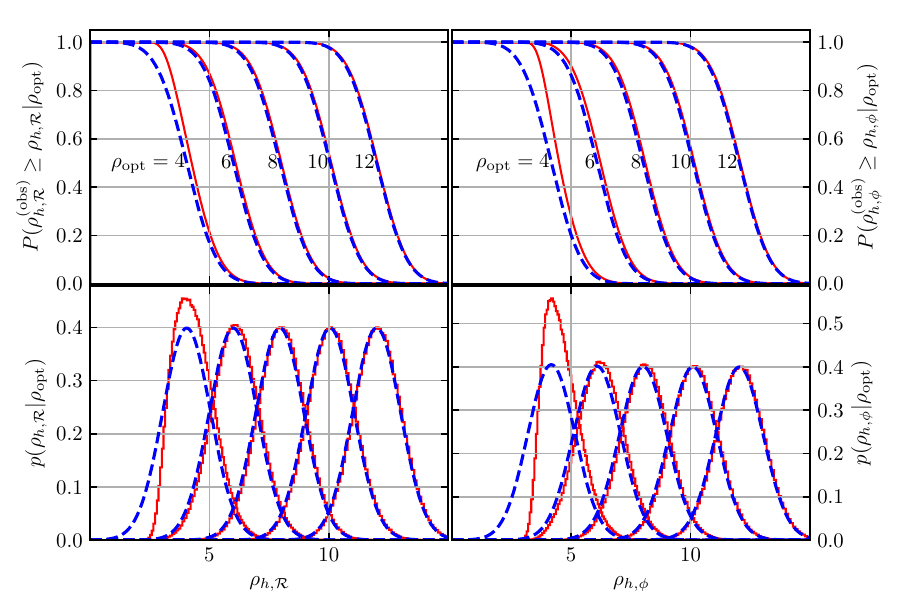}
    \end{center}
    \caption{
        Distributions of observed filter responses from a single detector maximized over the template bank (\emph{left}: $\rho_{h,\mathcal{R}}$ and \emph{right}: $\rho_{h,\phi}$) for several $\rho_\mathrm{opt}$.
        (\emph{red, solid}) Measured distributions from simulations and (\emph{blue, dashed}) predicted distributions of $\rho_\mathcal{R}$ and $\rho_\phi$ with the optimal template.
        Note that, for small $\rho_\mathrm{opt}$, the distributions can differ significantly.
        The agreement is much better for large $\rho_\mathrm{opt}$.
        Regardless, the tails to large $\rho_{h,\mathcal{R}}$ and $\rho_{h,\phi}$ always agree reasonably well.
        See Fig.~\ref{fig:money plot} for survival functions at fixed detection thresholds as a function of $\rho_\mathrm{opt}$.
    }
    \label{fig:distributions}
\end{figure*}

\section{Distributions of filter responses in stationary Gaussian noise}
\label{sec:filter response distributions}

Consider the distribution of both real and imaginary parts of the filter response from all templates within a bank simultaneously (not maximized over $\phi$ or other template parameters).
Specifically, consider the response of two templates ($h$ and $g$) in the presence of a true signal ($s$, which may or may not match either template) and stationary, additive Gaussian noise.
These may be summarized as follows
\begin{align}
    \mathrm{E}[\rho_\mathcal{R}(h)]
        & = \frac{1}{\mathcal{N}(h)} \left( 4\int df\, \frac{\mathcal{R}\{s^\ast h\}}{S} \right) \label{eq:mean snr real} \\
    \mathrm{E}[\rho_\mathcal{I}(h)]
        & = \frac{1}{\mathcal{N}(h)} \left( 4\int df\, \frac{\mathcal{I}\{s^\ast h\}}{S} \right) \\
    \mathrm{Cov}[\rho_\mathcal{R}(h), \rho_\mathcal{R}(g)]
        & = \mathrm{Cov}[\rho_\mathcal{I}(h), \rho_\mathcal{I}(g)] \\
        & = \frac{1}{\mathcal{N}(h)\mathcal{N}(g)} \left( 4\int df\, \frac{\mathcal{R}\{h^\ast g\}}{S} \right) \nonumber \\
    \mathrm{Cov}[\rho_\mathcal{R}(h), \rho_\mathcal{I}(g)]
        & = - \mathrm{Cov}[\rho_\mathcal{I}(h), \rho_\mathcal{R}(g)] \label{eq:cov snr real imag} \\
        & = \frac{1}{\mathcal{N}(h)\mathcal{N}(g)} \left( 4\int df\, \frac{\mathcal{I}\{h^\ast g\}}{S} \right) \nonumber
\end{align}
Furthermore, because $\rho_\mathcal{R}$ and $\rho_\mathcal{I}$ are linear in Gaussian data, they will also be Gaussian distributed.
One can therefore describe the entire response of the template bank in terms of a multivariate Gaussian distribution characterized by the moments defined in Eqs.~\ref{eq:mean snr real}-\ref{eq:cov snr real imag}.
It will also be convenient to define the optimal signal-to-noise ratio, obtained when the template exactly matches the true signal ($h=s$) in the absence of noise
\begin{equation}\label{eq:rho opt}
    \rho_\mathrm{opt}(s) = \frac{1}{\mathcal{N}(s)} \left( 4 \int df\,\frac{|s|^2}{S} \right) = \mathcal{N}(s)
\end{equation}
with the obvious extension to the network analog
\begin{equation}\label{eq:rho net opt}
    \rho_{\mathrm{net},\mathrm{opt}}(s) = \left(\sum\limits_i^{N_\mathrm{IFO}} \rho^2_{i,\mathrm{opt}}(s_i) \right)^{1/2}
\end{equation}
for a single signal as it appears in a network of detectors (i.e., accounting for the detectors' responses as a function of the binary's orientation and possibly frequency~\cite{Essick:2017}).
Additionally, by noting that
\begin{gather}
    \mathrm{Var}[\rho_\mathcal{R}(h)]
        = \mathrm{Var}[\rho_\mathcal{I}(h)] = 1 \\
    \mathrm{Cov}[\rho_\mathcal{R}(h), \rho_\mathcal{I}(h)] = 0
\end{gather}
it is readily apparent that $\rho_\phi(h)$ is non-central $\chi$-distributed\footnote{The non-central $\chi$-distribution describes the square root of the quadrature sum of standard Gaussian distributed variates. The number of Gaussian variates defines the number of degrees of freedom. I follow the same convention as \texttt{scipy.stats.ncx2}~\cite{scipy} for $\lambda$ (Eqs.~\ref{eq:lambda rho phi}-\ref{eq:lambda rho net phi}), although others effectively define the noncentrality parameters as $\lambda^2$.} with two degrees of freedom and non-centrality parameter
\begin{equation}\label{eq:lambda rho phi}
    \lambda_{\rho_\phi}(h) = \left( \mathrm{E}[\rho_\mathcal{R}(h)]^2 + \mathrm{E}[\rho_\mathcal{I}(h)]^2 \right)^{1/2} \leq \rho_\mathrm{opt}(h)
\end{equation}
with equality holding only when $h = s$.

Similarly, $\rho_{\mathrm{net},\mathcal{R}}(h)$ and $\rho_{\mathrm{net}, \phi}(h)$ are non-central $\chi$-distributed with $N_\mathrm{IFO}$ and $2 N_\mathrm{IFO}$ degrees of freedom, respectively, and non-centrality parameters
\begin{align}
    \lambda_{\rho_{\mathrm{net},\mathcal{R}}}(h) = \left( \sum\limits_i^{N_\mathrm{IFO}} \mathrm{E}[\rho_{i,\mathcal{R}}(h)]^2 \right)^{1/2} \label{eq:lambda rho net R}\\
    \lambda_{\rho_{\mathrm{net}, \phi}}(h) = \left( \sum\limits_i^{N_\mathrm{IFO}} \lambda^2_{\rho_{i,\phi}}(h) \right)^{1/2} \label{eq:lambda rho net phi}
\end{align}
When $h=s$, both of these non-centrality parameters are $\rho_{\mathrm{net},\mathrm{opt}}$.

Note that one can analytically describe the joint distribution of all the filter responses for many individual templates.
However, it is difficult, in general, to analytically describe the behavior of filter responses maximized over the template bank.
I instead numerically investigate the distribution of responses maximized over a template bank within the context of a simple waveform in Sec.~\ref{sec:sine-gaussian}.


\section{Filter responses maximized over a template bank}
\label{sec:sine-gaussian}

As Sec.~\ref{sec:filter response distributions} shows, the correlations between filter responses for different templates follow particular scalings with their expected values.
For example, the covariance between the response of a template $h$ and the optimal template (true signal) $s$ is
\begin{equation}
    \mathrm{Cov}[\rho_\mathcal{R}(h) \rho_\mathcal{R}(s)] = \frac{\mathrm{E}[\rho_\mathcal{R}(h)]}{\rho_\mathrm{opt}}
\end{equation}
Any template with $\mathrm{E}[\rho_\mathcal{R}] \sim \rho_\mathrm{opt}$ will be highly correlated with the response of the optimal template.
That is, the response of nearby templates will fluctuate up and down in lock-step with the response of the optimal template.
As such, maximizing the response over the template bank may not significantly affect the distribution of responses for loud signals.
The expected value for templates far from the true signal will be so much smaller than the optimal template's response that, even though their fluctuations are uncorrelated, it is very unlikely that the distant template's response can be scattered to large enough values to overcome its lower expected value.
As a result, the largest response in the bank for loud signals is very likely to be near the optimal template and therefore highly correlated with the optimal template's response.

Note that this behavior does not depend on the shapes of the templates, only on the similarity of nearby templates.
As such, I consider a simple model for which inner products (integrals over frequency) are analytically tractable, meaning the distribution of the real and imaginary filter responses for every template in the bank can be quickly computed.
With this in hand, it is straightforward to simulate the filter responses with different realizations of detector noise and in the presence of different signals.
I then investigate the behavior of the filter responses maximized over the template bank numerically.

Specifically, I define sine-Gaussian waveforms as follows
\begin{align}
    h(t) & = A \cos\left(2\pi \gamma (t - \tau) + \phi\right) e^{-(t - \tau)^2/2\sigma^2} \\
    h(f) & = \int dt\, e^{-2\pi i f t} h(t) \nonumber \\
        & = A \sqrt{2\pi} \sigma \cos\left(\phi\right) e^{-2\pi i \tau f - \pi^2\sigma^2(f-\gamma)^2}
\end{align}
\begin{widetext}
Assuming white noise ($S(f) = 1 \ \forall \ f$), I obtain
\begin{align}
    4\int df\,\mathcal{R}\{h^\ast g\} = \mathcal{A} \cos(\Psi) \quad \quad \quad \mathrm{and} \quad \quad \quad 
    4\int df\,\mathcal{I}\{h^\ast g\} = \mathcal{A} \sin(\Psi)
\end{align}
with
\begin{align}
    \mathcal{A} & = A_h A_g \left(\cos(\phi_h - \phi_g) + \cos(\phi_h + \phi_g)\right) \left(2\pi\frac{\sigma_h^2 \sigma_g^2}{\sigma_h^2 + \sigma_g^2}\right)^{1/2} \exp\left( -\frac{1}{2}\left( 4\pi^2 \frac{\sigma_h^2 \sigma_g^2}{\sigma_h^2 + \sigma_g^2}(\gamma_h-\gamma_g)^2 + \frac{(\tau_h - \tau_g)^2}{\sigma_h^2 + \sigma_g^2} \right) \right) \\
    \Psi & = 2\pi \left( \frac{\sigma_h^2 \gamma_h + \sigma_g^2 \gamma_g}{\sigma_h^2 + \sigma_g^2} \right) \left(\tau_h - \tau_g \right)
\end{align}
\end{widetext}
With these expressions, it is straightforward to construct the statistical properties of the response of an entire bank of sine-Gaussian templates to a signal from the same waveform family.

\begin{figure}[h]
    \begin{center}
        \Large{Single Detector Responses} \\
        \includegraphics[width=1.0\columnwidth]{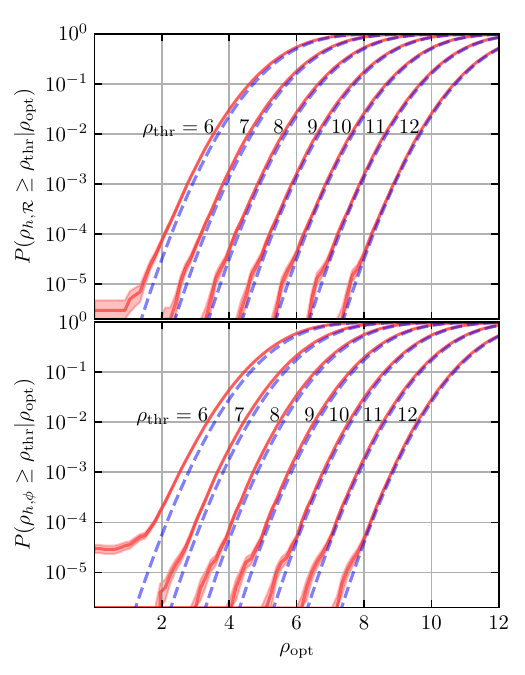}
    \end{center}
    \caption{
        Survival function of the response maximized over the template bank (\emph{top}: $\rho_{h,\mathcal{R}}$ and \emph{bottom}: $\rho_{h, \phi}$) as a function of $\rho_\mathrm{opt}$. 
        (\emph{red, solid}) Measured survival functions with (\emph{shaded}) $1\sigma$ uncertainty and (\emph{blue, dashed}) predicted survival functions for the optimal template.
    }
    \label{fig:money plot}
\end{figure}

I now examine how well one can approximate the distribution of responses maximized over a template bank with the distribution of the optimal template's response.
Sec.~\ref{sec:single IFO max over bank} examines the response of a single detector.
Sec.~\ref{sec:network max over bank} extends this to networks of multiple detectors.
I assume the true signal is contained within the bank and that the template bank is dense enough that there will always be a good match to the true signal.
What follows can be extended to cases where the bank does not contain a close match to the true signal, although one would need to replace $\rho_\mathrm{opt}$ with $\mathrm{E}[\rho_\mathcal{R}]$ for the template that best matches the signal.


\begin{figure*}
    \begin{center}
        \Large{2-Detector Network Responses} \\
    \end{center}
    \begin{minipage}{1.0\columnwidth}
        \begin{center}
            $\rho_{1,\mathrm{opt}} = \rho_{\mathrm{net},\mathrm{opt}}$ and $\rho_{2,\mathrm{opt}} = 0$ \\
            \includegraphics[width=1.0\textwidth]{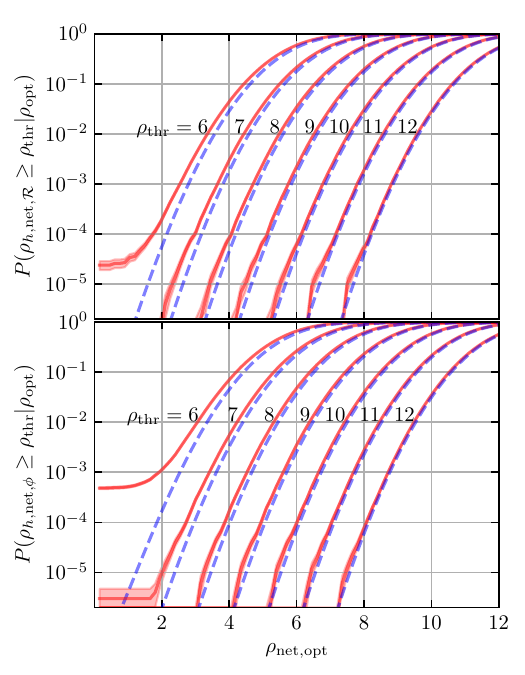}
        \end{center}
    \end{minipage}
    \hfill
    \begin{minipage}{1.0\columnwidth}
        \begin{center}
            $\rho_{1,\mathrm{opt}} = \rho_{2,\mathrm{opt}} = \rho_{\mathrm{net},\mathrm{opt}} / \sqrt{2}$ \\
            \includegraphics[width=1.0\textwidth]{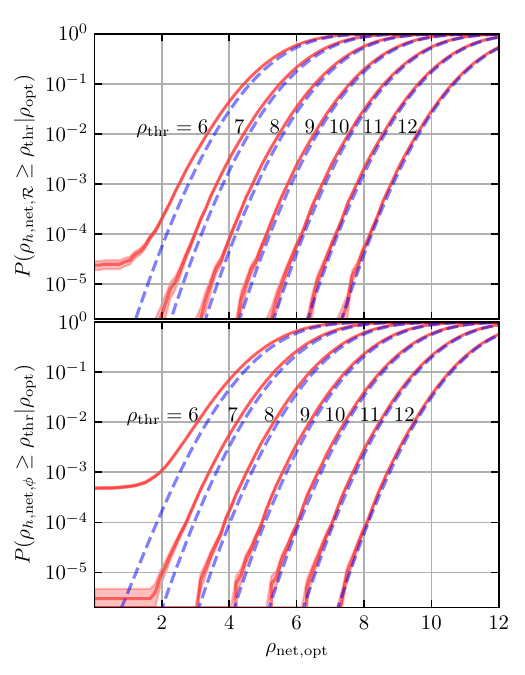}
        \end{center}
    \end{minipage}
    \caption{
        Survival functions at a fixed detection threshold as a function of $\rho_{\mathrm{net},\mathrm{opt}}$ for a two-detector network with (\emph{left}) all signal power in a single detector and (\emph{right}) equal signal power in each detector.
        Colors and line-styles are the same as Fig.~\ref{fig:money plot}.
        As expected, the survival functions are larger at each value of $\rho_{\mathrm{net},\mathrm{opt}}$ for multi-detector networks compared to single detector networks, particularly when both $\rho_\mathrm{thr}$ and $\rho_\mathrm{opt}$ are small.
        However, there is no significant dependence on the balance of signal power between detectors.
        Compare to Figs.~\ref{fig:money plot} and~\ref{fig:3ifo}.
    }
    \label{fig:2ifo}
\end{figure*}

\begin{figure}
    \begin{center}
        {\Large{3-Detector Network Responses}} \\
        $\rho_{i,\mathrm{opt}} = \rho_{\mathrm{net},\mathrm{opt}}/\sqrt{3} \ \forall \ i$ \\
        \includegraphics[width=1.0\columnwidth]{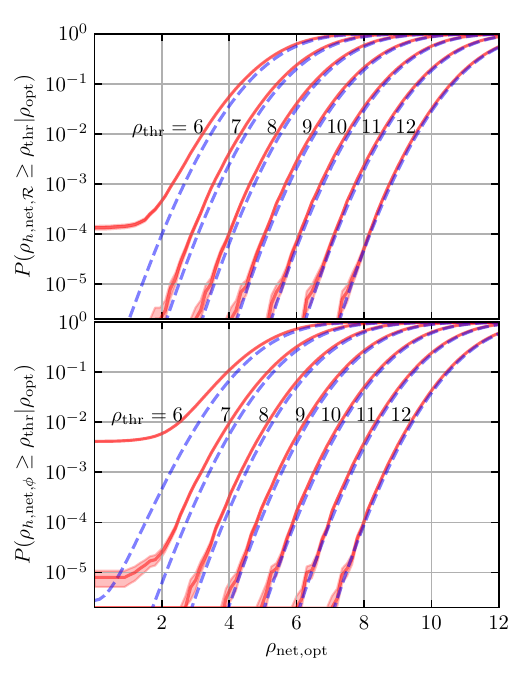}
    \end{center}
    \caption{
        Survival functions at fixed detection thresholds as a function of $\rho_{\mathrm{net},\mathrm{opt}}$ for a three-detector network assuming all detectors have equal individual optimal signal-to-noise ratios.
        Colors and line-styles are the same as Fig.~\ref{fig:money plot}.
        Compare to Figs.~\ref{fig:money plot} and~\ref{fig:2ifo}.
    }
    \label{fig:3ifo}
\end{figure}

\subsection{Response of a single detector}
\label{sec:single IFO max over bank}

Fig.~\ref{fig:distributions} shows empirical distributions of $\rho_{h,\mathcal{R}}$ and $\rho_{h,\phi}$ for a single detector as a function of $\rho_\mathrm{opt}$.\footnote{Figs.~\ref{fig:distributions}-\ref{fig:3ifo} use $21^3=9261$ templates evenly spaced between $\tau\in[-5, +5]$, $\gamma\in[0, 10]$, $\sigma\in[0.1,2.0]$, and $\phi=0$. They consider an injection with $\tau=0$, $\gamma=5$, $\sigma=1$, and $\phi=0$.}
As expected, the relative importance of maximizing over the template bank is smaller for larger $\rho_\mathrm{opt}$.
Nevertheless, there is always reasonably good agreement between the optimal template's distribution ($\rho_\mathcal{R}$ and $\rho_\phi$) and the distribution of responses maximized over the bank ($\rho_{h,\mathcal{R}}$ and $\rho_{h,\phi}$) in the tails to large $\rho$.
Fig.~\ref{fig:money plot} shows the survival function of different filter responses maximized over the bank as a function of $\rho_\mathrm{opt}$ for several detection thresholds ($\rho_\mathrm{thr}$).
There is good agreement between the observed distribution of filter responses maximized over the template bank and the simple analytic model for the optimal template's response.
Even for relatively low thresholds ($\rho_\mathrm{thr}=6$), the observed survival function does not differ by more than a factor of $\sim 2$ from the approximation based on the optimal template until very quiet signals ($\rho_\mathrm{opt} \lesssim 2$).

While the exact behavior of the response maximized over the bank may depend on the size of the bank (number of effectively independent templates), one can accurately approximate the distribution of, e.g., $\rho_{h, \phi}$ with the distribution of the optimal template's $\rho_\phi$ at realistic detection thresholds ($\rho_\mathrm{thr} \gtrsim 9$, see Sec.~\ref{sec:real injections}).
Therefore, one can reasonably approximate the fraction of signals with low $\rho_\mathrm{opt}$ that are scattered above realistic detection thresholds (detection probability) using simple analytic models of the response of the optimal template without directly maximizing the response over the template bank.


\subsection{Response from a network of detectors}
\label{sec:network max over bank}

The distributions of network statistics differ from the single-detector study in Sec.~\ref{sec:single IFO max over bank}.
Specifically, assuming fixed $\rho_{\mathrm{net}, \mathrm{opt}}$, each detector in a network will have smaller $\rho_{i,\mathrm{opt}}$ and therefore be more subject to noise fluctuations.
That is, the number of degrees of freedom in the non-central $\chi$-distribution matters.

As implicit within Eqs.~\ref{eq:snr h net R} and~\ref{eq:snr h net phi}, it is common practice to consider the response of identical templates in all detectors within the network.\footnote{The transfer function between astrophysical signals and detector outputs is often modeled as a multiplicative constant for each detector (see, e.g.,~\citet{Essick:2017}). As such, the spectral shape of the signal is expected to be identical in all detectors; only the overall amplitude and phase may differ.}
I therefore consider the response of the same template within each detector in the network.
I further assume noise realizations for each detector are independent but the responses for different templates within each detector are correlated.
Finally, I maximize the network response over the template bank.

Figs.~\ref{fig:2ifo} and~\ref{fig:3ifo} show the survival function of network responses ($\rho_{h,\mathrm{net},\mathcal{R}}$ and $\rho_{h,\mathrm{net},\phi}$) as a function of $\rho_{\mathrm{net},\mathrm{opt}}$.
Fig.~\ref{fig:2ifo} considers a two-detector network and Fig.~\ref{fig:3ifo} considers a three-detector network.

As expected, the survival functions are larger with larger networks, and this impacts $\rho_{h,\mathrm{net},\phi}$ more than $\rho_{h,\mathrm{net},\mathcal{R}}$.
That is expected, as maximization over $\phi$ will only increase the observed response.
Put another way, the network search statistic has more degrees of freedom than a single-detector response, and there are more ways that noise can broaden the distribution.
This is particularly evident in the plateaus at small $\rho_{\mathrm{net}, \mathrm{opt}}$.
Higher plateaus in multi-detector networks show that there is a higher chance of noise alone producing observed responses above the the detection threshold.\footnote{See, e.g.,~\citet{Morras:2022} for more discussion and numerical experiments using realistic PSDs and GW waveforms. They only consider correlations between the response of a single template at nearby times and estimate the false alarm rate (FAR). I consider correlations among all template parameters, including time, but do not estimate the FAR.}

Interestingly, there is no strong dependence on the balance of signal power between detectors.
Fig.~\ref{fig:2ifo} shows that the observed survival functions are nearly identical when the signal power is equally divided between detectors and when it is entirely within a single detector.

I therefore conclude that one can also model the distribution of filter responses from a network of detectors that are maximized over a template bank reasonably well with only the distribution of the optimal response if the detection threshold is reasonably large ($\rho_\mathrm{thr} \gtrsim 9$) regardless of the distribution of signal power between detectors.
Again, this approximation improves as either $\rho_\mathrm{thr}$ or $\rho_{\mathrm{net},\mathrm{opt}}$ increases.

In reality, many searches' detection statistics may act somewhere between $\rho_{\mathrm{net},\mathcal{R}}$ and $\rho_{\mathrm{net},\phi}$, as they may maximize separately in each detector but additionally place requirements on the balance of power between detectors and the consistency of $\phi$ estimated from different detectors.
In this sense, estimating $P(\rho_\mathrm{obs} \geq \rho_\mathrm{thr}|\rho_\mathrm{opt})$ based on $\rho_{\mathrm{net},\mathcal{R}}$ may be pessimistic (fewer signals will be scattered above $\rho_\mathrm{thr}$ than in reality) while estimating it based on $\rho_{\mathrm{net},\phi}$ may be optimistic (more signals will be scattered above $\rho_\mathrm{thr}$ than in reality).


\begin{figure}[b]
    \includegraphics[width=1.0\columnwidth]{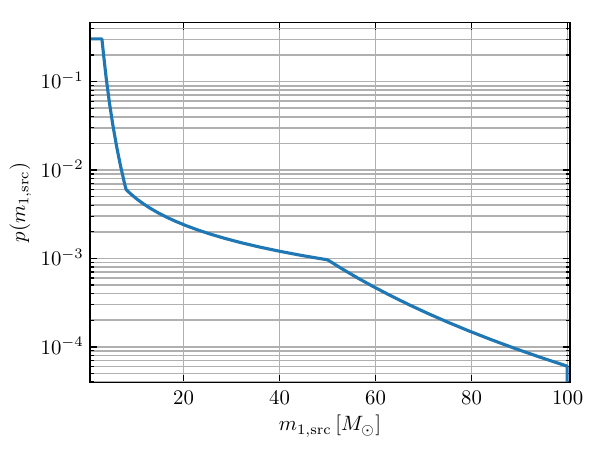}
    \caption{
        The distribution of primary masses assumed within Sec.~\ref{sec:real injections}.
        This distribution captures the key features identified in~\citet{Farah:2022} and~\citet{GWTC-3-RnP}: a high rate of low-mass systems followed by a sharp drop-off to a relatively flat distribution until a gentle steepening at the highest masses.
    }
    \label{fig:reference distribution mass1_source}
\end{figure}

\section{Comparison to real injections from O3}
\label{sec:real injections}

I now turn to the question of how accurately semianalytic sensitivity estimates can emulate the distribution of signals detected by real searches.
To that end, I generated a large number of semianalytic injections using the~\texttt{IMRPhenomXPHM} compact binary approximant~\cite{Pratten:2020} and reference PSDs for O3\footnote{Semianalytic injections were generated using the representative O3 PSDs for LIGO Hanford, LIGO Livingston, and Virgo that were used to compute the hopeless signal-to-noise ratio cut for the real O3 injections~\cite{GWTC-3-injections}. Note, however, that some of the signal-to-noise ratios reported in that release are only based on the LIGO detectors, and they are all estimates of $\rho_{\mathrm{net},\mathrm{opt}}$. See Fig.~\ref{fig:psd}.}.
The semianalytic injections are selected based on a threshold on $\rho_{\mathrm{net},\phi}$, which itself is a random variate drawn from a non-central $\chi$ distribution given $\rho_{\mathrm{net},\mathrm{opt}}$ as computed with the reference PSDs.
For simplicity, they do not account for PSD variability and instead assume a fixed PSD for each detector.
I then compare them to the publicly available injections~\cite{GWTC-3-injections}.
To compare the real and semianalytic distributions, I first reweigh both the semianalytic and real injections to follow the same astrophysical distribution
\begin{itemize}
    \item a four-piece broken powerlaw between $1$--$100\,M_\odot$ for the source-frame primary mass (Fig.~\ref{fig:reference distribution mass1_source}),
    \item a power law for the source-frame secondary mass with a minimum of $1\,M_\odot$ and an exponent of 1.0,
    \item independent and identically distributed component spins that are isotropic in direction and uniform in magnitude (up to a maximum dimensionless spin of 0.4), and
    \item a redshift distribution corresponding to a local merger rate that grows as $(1+z)$ (Eq.~\ref{eq:redshift distribution} with $\kappa=1.0$).
\end{itemize}

\begin{figure}
    \includegraphics[width=1.0\columnwidth, clip=True, trim=0.0cm 0.0cm 0.0cm 1.0cm]{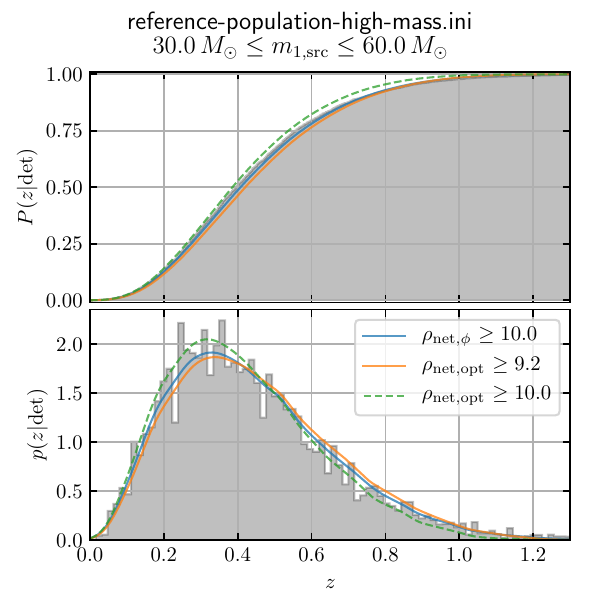}
    \caption{
        Distributions of detected redshifts ($z$) assuming a reference astrophysical population with the additional requirement that $30\,M_\odot \leq m_{1,\mathrm{src}} \leq 60\,M_\odot$.
        (\emph{grey}) Real injections selected based on GstLAL's FAR ($\leq 1/\mathrm{year}$), (\emph{solid blue}) semianalytic selection based on $\rho_{\mathrm{net},\phi}$ with optimized $\rho_\mathrm{thr}$, (\emph{solid orange}) semianalytic selection based on $\rho_{\mathrm{net},\mathrm{opt}}$ with optimized $\rho_\mathrm{thr}$, and (\emph{dashed green}) semianalytic selection based on $\rho_{\mathrm{net},\mathrm{opt}}$ with the same $\rho_\mathrm{thr}$ as $\rho_{\mathrm{net},\phi}$.
    }
    \label{fig:detected distribution}
\end{figure}

\begin{figure}
    \includegraphics[width=1.0\columnwidth]{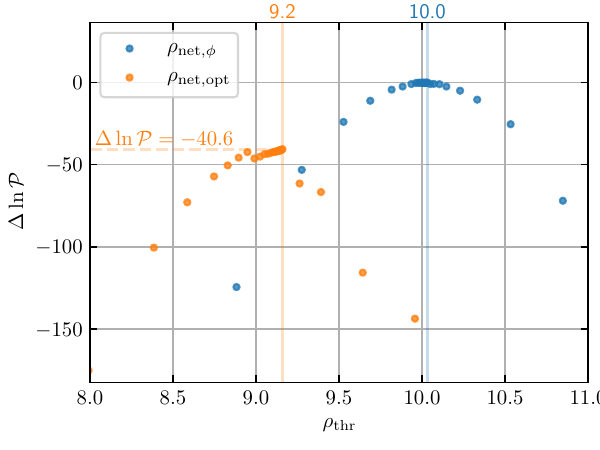}
    \caption{
        Differences in the cross-entropy (Eq.~\ref{eq:cross entropy}) between the real distribution of detected redshifts in Fig.~\ref{fig:detected distribution} ($30\,M_\odot \leq m_{1,\mathrm{src}} \leq 60\,M_\odot$) and semianalytic models based on (\emph{blue}) $\rho_{\mathrm{net},\phi}$ and (\emph{orange}) $\rho_{\mathrm{net},\mathrm{opt}}$ for various $\rho_\mathrm{thr}$.
    }
    \label{fig:cross-entropy}
\end{figure}

\begin{figure}
    \begin{center}
        \includegraphics[width=1.0\columnwidth, clip=True, trim=0.3cm 0.2cm 0.4cm 0.3cm]{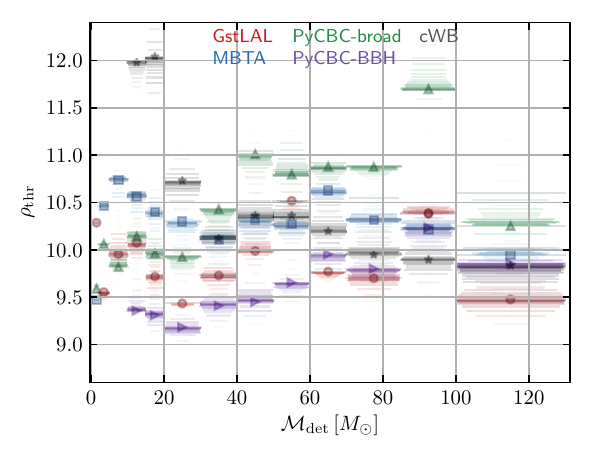}
    \end{center}
    \caption{
        Mass-dependence of the optimized $\rho_\mathrm{thr}$ for models based on $\rho_{\mathrm{net},\phi}$ as a function of detector-frame chirp mass ($\mathcal{M}_\mathrm{det}$) for (\emph{red circles}) GstLAL~\cite{Messick:2017, Sachdev:2019, Hanna:2019, Cannon:2021}, (\emph{blue squares}) MBTA~\cite{Adams:2015, Aubin:2020}, (\emph{green upward triangles}) PyCBC-broad and (\emph{purple sideways triangles}) PyCBC-BBH~\cite{Allen:2004, DalCanton:2014, Usman:2015, Nitz:2017, Davies:2020, Allen:2005}, and (\emph{grey stars}) cWB~\cite{Klimenko:2011, Klimenko:2004, Klimenko:2015}.
        Shading approximates the precision of the fit to real injections as a function of $\rho_\mathrm{thr}$ (see Fig.~\ref{fig:cross-entropy}).
    }
    \label{fig:threshold by mass}
\end{figure}

With large sets of semianalytic injections, I identify the preferred threshold $\rho_\mathrm{thr}$ by extremizing the cross-entropy ($\ln \mathcal{P}$) between the real detected injections and a density estimate ($\hat{p}$) derived from the semianalytic injections
\begin{equation}\label{eq:cross entropy}
    \ln \mathcal{P} = \sum_i^{N} \ln \hat{p}(\theta_i|\rho \geq \rho_\mathrm{thr}, \Lambda)
\end{equation}
where the samples $\theta_i$ are drawn from the detected distribution defined by a false alarm rate (FAR) threshold with real injections
\begin{equation}\label{eq:that one}
    \theta_i \sim p(\theta_i|\mathrm{FAR} \leq \mathrm{FAR}_\mathrm{thr}, \Lambda)
\end{equation}
I denote single-event parameters (e.g., masses, spins, redshift, etc.) for the $i^\mathrm{th}$ event with $\theta_i$ and population parameters (e.g., minimum and maximum mass, etc) as $\Lambda$.
This procedure is equivalent to minimizing the Kullback-Leibler divergence between the model derived from the semianalytic samples and the distribution from which the real detected samples are drawn.
In what follows, I only optimize $\rho_\mathrm{thr}$ based on the distribution of detected redshifts ($\theta = z$ in Eqs.~\ref{eq:cross entropy} and~\ref{eq:that one}).
There is no reason why this optimization could not be extended to an arbitrary set of single-event parameters, though.
I only use the redshift for computational simplicity and because it produces accurate emulators.

\begin{figure*}
    \begin{center}
        \includegraphics[width=0.95\textwidth]{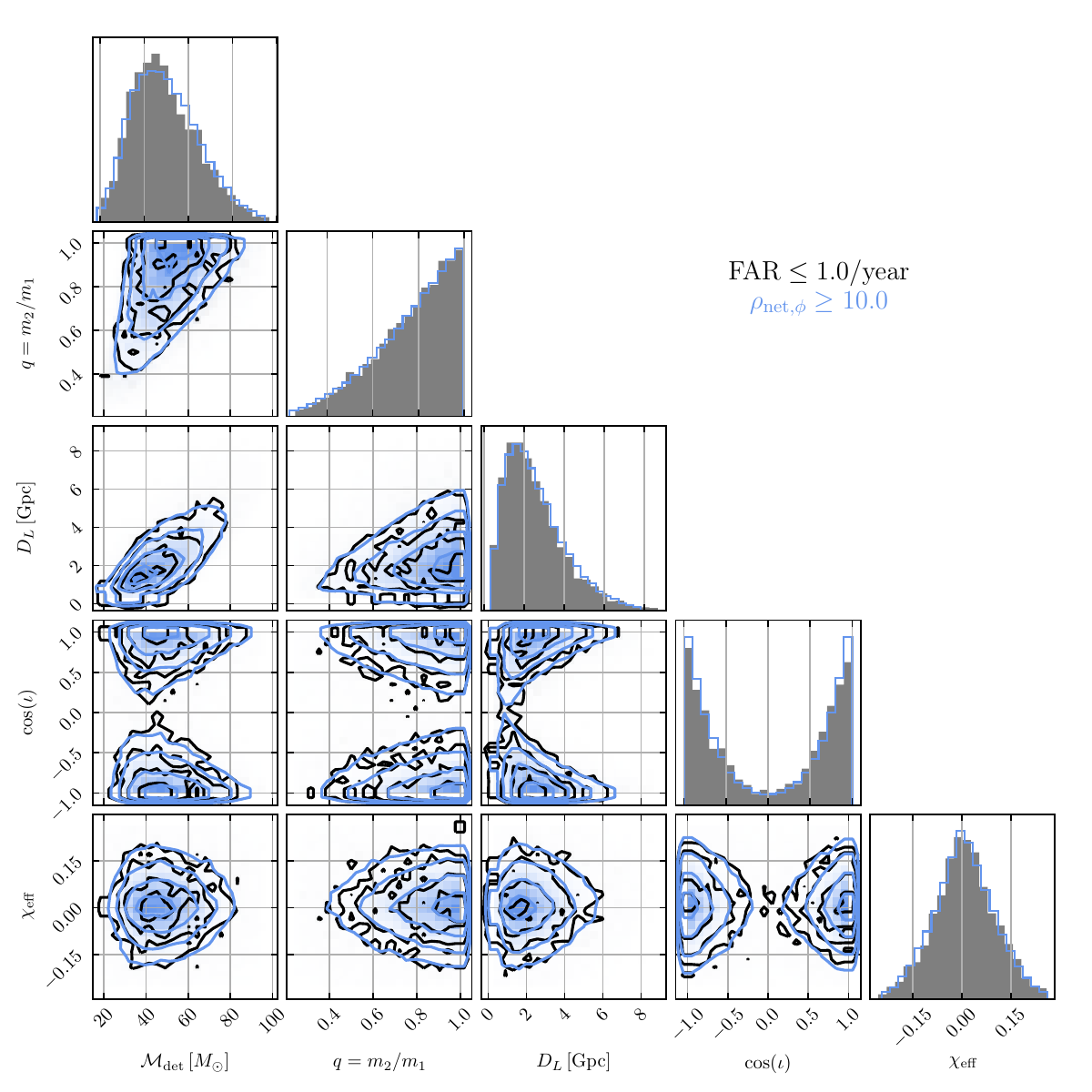}
    \end{center}
    \caption{
        Detected distributions of additional single-event parameters for events with $30\,M_\odot \leq m_{1,\mathrm{src}} \leq 60\,M_\odot$ with (\emph{black}) real injections selected based on GstLAL's FAR ($\leq 1/\mathrm{year}$) and (\emph{blue}) an optimized semianalytic model with a selection based on $\rho_{\mathrm{net},\phi}$ ($\geq 10.0$).
    }
    \label{fig:detected corner}
\end{figure*}

There is only a single free parameter in this procedure: $\rho_\mathrm{thr}$.
The rest of the dependence on the binary parameters is implicitly defined by the detector sensitivities (PSDs) and the physics of compact binary coalescences.
Fig.~\ref{fig:detected distribution} demonstrates general trends which are seen for all matched-filter searches included in the O3 injections~\cite{GWTC-3-injections}.
It shows the distribution of true redshifts of detected systems for the subset of the astrophysical population with source-frame primary masses between $30$--$60\,M_\odot$.
Real injections are selected based on a threshold on the FAR assigned by GstLAL~\cite{Messick:2017, Sachdev:2019, Hanna:2019, Cannon:2021} ($\leq 1/\mathrm{year}$ as in~\citet{GWTC-3-RnP}) whereas semianalytic injections are drawn separately and selected based on either $\rho_{\mathrm{net},\phi}$ or $\rho_{\mathrm{net},\mathrm{opt}}$, respectively.
Again, the only information about the detector noise included within the semianalytic models are the reference PSDs from O3.

While both semianalytic estimates can produce reasonable fits to the data, I find that selecting based on the observed signal-to-noise ratio ($\rho_{\mathrm{net},\phi}$) almost always produces a better fit to the data than selecting on the optimal signal-to-noise ratio ($\rho_{\mathrm{net},\mathrm{opt}})$.

Briefly, this is likely due to the fact that the distribution based on $\rho_{\mathrm{net},\phi}$ is intrinsically wider than the distribution based on $\rho_{\mathrm{net},\mathrm{opt}}$, regardless of $\rho_\mathrm{thr}$.
The extra variance is important when reproducing the real detected distribution; models based on $\rho_{\mathrm{net},\mathrm{opt}}$ produce distributions that are slightly too narrow.
Typically, I find optimized $\rho_\mathrm{thr}$ for $\rho_{\mathrm{net},\mathrm{opt}}$ that are smaller than the equivalent threshold for $\rho_{\mathrm{net},\phi}$, which suggests that the optimization favors matching the tail to larger redshifts over the (relatively few) nearby sources.
This is also apparent in Fig.~\ref{fig:detected distribution} since the detected distribution based on $\rho_{\mathrm{net},\mathrm{opt}}$ with an optimized $\rho_\mathrm{thr}$ is shifted to larger redshifts (underestimates the detection probability at low redshifts).
Furthermore, if we use $\rho_\mathrm{thr}$ optimized for $\rho_{\mathrm{net},\phi}$ but apply it to $\rho_{\mathrm{net},\mathrm{opt}}$, we see that the distribution is shifted to lower redshifts (underestimates the detection probability at high redshifts).\footnote{It may be tempting to estimate $\rho_\mathrm{thr}$ based on the smallest $\rho_{\mathrm{net},\phi}$ reported within a catalog, and then apply that within a semianalytic model. Fig.~\ref{fig:detected distribution} shows that this could easily lead to underestimates of the detection probability at high redshifts if that model is based on $\rho_{\mathrm{net},\mathrm{opt}}$ instead of $\rho_{\mathrm{net},\phi}$. This is also apparent in Fig.~\ref{fig:cross-entropy}, where one can see that applying $\rho_\mathrm{thr}$ optimized for $\rho_{\mathrm{net},\phi}$ ($\rho_\mathrm{thr} = 10$) to $\rho_{\mathrm{net},\mathrm{opt}}$ produces a dramatically worse fit ($\Delta \ln \mathcal{P} \sim -150$).}
The distribution based on $\rho_{\mathrm{net},\phi}$ falls between these two and matches the detected distribution well at all redshifts.

Fig.~\ref{fig:cross-entropy} quantifies how much better the model based on $\rho_{\mathrm{net},\phi}$ fits the real injections compared to the model based on $\rho_{\mathrm{net}, \mathrm{opt}}$.
Here, I show the cross entropy (Eq.~\ref{eq:cross entropy}) as a function of $\rho_\mathrm{thr}$.
The behavior for $\rho_{\mathrm{net},\phi}$ shows a clear local maximum, with good matches over a relatively small range of $\rho_\mathrm{thr}$.
This suggests that it is able to model the true detected distribution well.
Conversely, the behavior for $\rho_{\mathrm{net},\mathrm{opt}}$ is more complex, including sudden changes in the cross entropy and multimodality.
This suggests that no single value of $\rho_\mathrm{thr}$ is able to match the true detected distribution well.
The cross-entropy is also significantly lower overall for the $\rho_{\mathrm{net}, \mathrm{opt}}$ model.

Building upon this, Fig.~\ref{fig:detected corner} shows the distributions of other single-event parameters using real injections and the semianalytic approximation based on $\rho_{\mathrm{net},\phi}$.
I stress that the value of $\rho_\mathrm{thr}$ was chosen by matching only the detected distribution of redshifts.
Nevertheless, I find consistent distributions for the all other single-event parameters.
Table~\ref{tab:o4 projections} confirms this by comparing the fractions of the detected population within different mass ranges when selection is performed with real injections and with a semianalytic model based on $\rho_{\mathrm{net},\phi}$.
They match to within the uncertainty from the finite number of injections almost everywhere.
This further suggests that selecting based on $\rho_{\mathrm{net},\phi}$ concisely captures much of the relevant physics that drives search sensitivity.

Finally, there is one further trend that is worth mentioning.
Figs.~\ref{fig:detected distribution}-\ref{fig:detected corner} consider a subpopulation with relatively large primary masses ($30\,M_\odot \leq m_{1,\mathrm{src}} \leq 60\,M_\odot$) and optimize $\rho_\mathrm{thr}$ to match the real detected distribution obtained by thresholding on only a single search's FAR (GstLAL~\cite{Messick:2017, Sachdev:2019, Hanna:2019, Cannon:2021}).
While I find similar behavior for other searches and mass ranges, the optimal threshold depends on both the search and the component masses.
Nevertheless, there is a clear trend toward larger $\rho_\mathrm{thr}$ at higher masses for most searches between $\mathcal{M}_\mathrm{det}\sim 20$--$60\,M_\odot$.
That is, high-mass signals must be louder to be detected.
Fig.~\ref{fig:threshold by mass} summarizes this trend for several searches; PyCBC-broad~\cite{Allen:2004, DalCanton:2014, Usman:2015, Nitz:2017, Davies:2020, Allen:2005} shows the most dramatic dependence.
Interestingly, even cWB~\cite{Klimenko:2011, Klimenko:2004, Klimenko:2015} shows similar behavior within this mass range even though that search is not based upon a matched filter.
Overall, though, the mass-dependence is mild, and the shift in $\rho_\mathrm{thr}$ across the entire mass range is of comparable size to the standard deviation of $\rho_{\mathrm{net},\phi}$ for a three-detector network.

As such, it is reasonable to expect semianalytic estimate to provide a reasonable approximation to the true detected distribution even with a single $\rho_\mathrm{thr}$ applied across the entire mass range.
Furthermore, semianalytic estimates with this $\rho_\mathrm{thr}$ should trivially generalize to other detector networks or PSDs.


\begin{figure*}
    \begin{minipage}{0.49\textwidth}
        \includegraphics[width=1.0\textwidth]{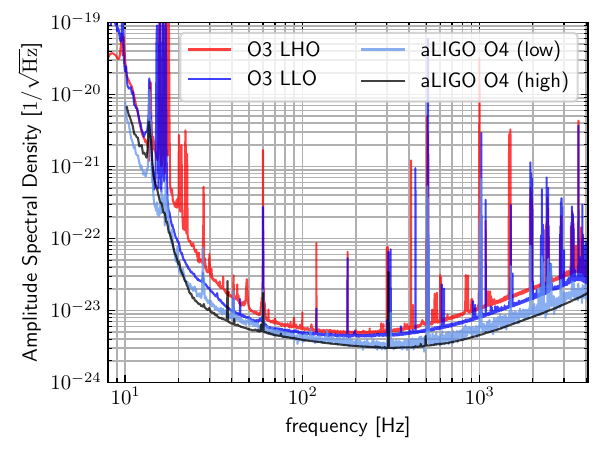}
    \end{minipage}
    \begin{minipage}{0.49\textwidth}
        \includegraphics[width=1.0\textwidth]{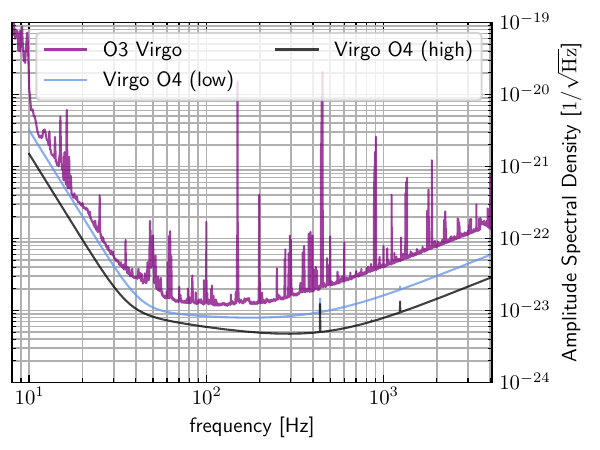}
    \end{minipage}
    \caption{
        Representative PSDs used to generate the semianalytic sensitivity estimates used in this work.
        O3 PSDs are the same ones that were used to generate hopeless signal-to-noise ratio cuts in Ref.~\cite{GWTC-3-injections}.
        Projections for low and high O4 sensitivity are from Ref.~\cite{o4-psd}.
        As of the time of writing, Virgo has not yet joined O4 and the LIGOs have low-frequency sensitivities are closer to their O3 values than the projections for O4.
    }
    \label{fig:psd}
\end{figure*}

\subsection{Projections for O4}
\label{sec:projections for O4}

Given projected PSDs for O4~\cite{o4-psd}, I construct semianalytic sensitivity estimates assuming a three-detector network (HLV) and a mass-independent threshold on the observed phase-maximized signal to noise ratio: $\rho_{\mathrm{net},\phi} \geq 10$.
I adopt the same reference distribution for source-frame component masses, spins, and redshift as before.
Table~\ref{tab:o4 projections} reports the expected fractions of the detected population that fall within several categories based on the source-frame component masses.
These include those defined by the LVK and estimated in low-latency as part of the public alert system~\cite{lvk-userguide}, but I provide additional categories as well.

Table~\ref{tab:o4 projections} shows that the fractions of the detected distribution estimated with a semianalytic model with mass-independent $\rho_\mathrm{thr}$ and (fixed) reference PSDs nearly match the fractions computed from real injections to within the uncertainty from the finite sample size.
Additionally, projections for O4 suggest that the fraction of each category in the detected distribution may change, but they will not be radically altered.
Therefore, one should still expect there to be far more binary black hole (BBH) mergers detected during O4 than mergers containing a low-mass object.

In addition to ratios within the detected distribution, it may be helpful to estimate the total number of detected events that should be expected during O4 based on the number of events previously observed.
To wit, hierarchical Bayesian models of GW populations assume an inhomogeneous Poisson process with differential rate density parametrized by an overall rate ($R$) and population parameters ($\Lambda$)
\begin{equation}
    \frac{dN}{d\theta} = R p(\theta|\Lambda)
\end{equation}
From this, the expected number of events within an observing run can be calculated via from the time-averaged detector sensitivity (Eq.~\ref{eq:average detection}) as
\begin{equation}
    \mathrm{E}[N|\Lambda] = \mu(\Lambda) = T R P(\mathrm{det}|\Lambda)
\end{equation}
where $T$ denotes the length of the run.
Even without knowledge of the overall rate, we can construct the following relation
\begin{equation}
    C \equiv \frac{\mu_\mathrm{new}}{\mu_\mathrm{old}} = \left(\frac{T_\mathrm{new}}{T_\mathrm{old} }\right) \left(\frac{P_\mathrm{new}(\mathrm{det}|\Lambda)}{P_\mathrm{old}(\mathrm{det}|\Lambda)}\right)
\end{equation}
and thereby estimate the expected number of new detections given the relative sensitivity of the new run and the old run to the same population.
However, $\mu$ is not observed directly, but instead one observes a number drawn from a Poisson distribution conditioned on $\mu$
\begin{equation}
    N \sim \frac{\mu^N}{N!} e^{-\mu}
\end{equation}
Assuming a flat prior for $\mu_\mathrm{old}$ so that $p(\mu_\mathrm{old}|N_\mathrm{old}) \propto p(N_\mathrm{old}|\mu_\mathrm{old})$, we obtain
\begin{align}
    P(N_\mathrm{new}|N_\mathrm{old})
        & = \int d\mu_\mathrm{old}\, p(\mu_\mathrm{old}|N_\mathrm{old}) P(N_\mathrm{new}|\mu_\mathrm{new} = C \mu_\mathrm{old}) \nonumber \\
        & = \frac{C^{N_\mathrm{new}}}{(1+C)^{N_\mathrm{new} + N_\mathrm{old} + 1}} \left(\begin{matrix} N_\mathrm{new} + N_\mathrm{old} \\ N_\mathrm{old} \end{matrix} \right)
\end{align}
Using semianalytic estimates for $P(\mathrm{det}|\Lambda)$ with representative PSDs for O3, projected PSDs for O4, and selection via $\rho_{\mathrm{net},\phi} \geq 10$, I find \result{$P_\mathrm{O4}(\mathrm{det}|\Lambda)/P_\mathrm{O3}(\mathrm{det}|\Lambda) = 3.439 \pm 0.006$ ($2.227 \pm 0.004$) given the high (low) sensitivity predicted for O4}.
Assuming O4 lasts 18 months and given $N_\mathrm{O3} = 63$ confident detections (FAR $\leq 1/\mathrm{year}$) after one year at O3 sensitivity, I predict \result{$N_\mathrm{O4} = 324^{+80}_{-69}$ ($210^{+54}_{-47}$) new confident detections during O4 with high (low) sensitivity}.
This correspond to approximately \result{$4.1^{+1.0}_{-0.9}$ ($2.7^{+0.7}_{-0.6}$) detections/week}.\footnote{See Ref.~\cite{lvk-userguide} for other predictions assuming a different astrophysical model and selection threshold.}

\begin{turnpage}

\begin{table*}
    \caption{
        Expected fractions of the detected population within different categories of events for real O3 injections (selected with GstLAL's FAR $\leq 1/\mathrm{year}$) as well as semianalytic injections ($\rho_{\mathrm{net},\phi} \geq 10$) assuming three detector networks (HLV) with fixed PSDs for O3~\cite{GWTC-3-injections} and two projected sensitivities for O4~\cite{o4-psd} (see Fig.~\ref{fig:psd}).
        I provide several categories, including those defined by the LVK~\cite{lvk-userguide}.
        Reported errors approximate 1$\sigma$ uncertainty from the finite number of injections.
    }
    \label{tab:o4 projections}
    {\renewcommand{\arraystretch}{1.4}
    \begin{tabular}{@{\extracolsep{0.1cm}}l c c r r r r}
        \hline
        \hline
        \multicolumn{3}{c}{\multirow{3}{*}{Classification}} & \multicolumn{4}{c}{Fraction of Detected Population} \\
         & & & \multicolumn{2}{c}{O3} & \multicolumn{2}{c}{O4} \\
        \cline{4-7}
         & & & \multicolumn{1}{c}{FAR} & \multicolumn{1}{c}{$\rho_{\mathrm{net},\phi}$} & \multicolumn{1}{c}{low} & \multicolumn{1}{c}{high} \\
        \hline
        \multirow{5}{*}{LVK}
         & HasMassGap & ($3\,M_\odot \leq m_{2,\mathrm{src}} \leq 5\,M_\odot$) OR ($3\,M_\odot \leq m_{1,\mathrm{src}} < 5\,M_\odot$) &
            \result{\oThreeRealHasMassGap} & \result{\oThreeSemiHasMassGap} & \result{\oFourLowHasMassGap} & \result{\oFourHighHasMassGap} \\
         & HasNS & $M_\odot < m_{2,\mathrm{src}} \leq 3\,M_\odot$ &
            \result{\oThreeRealHasNS} & \result{\oThreeSemiHasNS} & \result{\oFourLowHasNS} & \result{\oFourHighHasNS} \\
        \cline{2-7}
         & BNS & $M_\odot < m_{2,\mathrm{src}} \leq m_{1,\mathrm{src}} \leq 3\,M_\odot$ &
            \result{\oThreeRealBNS} & \result{\oThreeSemiBNS} & \result{\oFourLowBNS} & \result{\oFourHighBNS} \\
         & NSBH & $M_\odot < m_{2,\mathrm{src}} \leq 3\,M_\odot < m_{1,\mathrm{src}} < 100\,M_\odot$ &
            \result{\oThreeRealNSBH} & \result{\oThreeSemiNSBH} & \result{\oFourLowNSBH} & \result{\oFourHighNSBH} \\
         & BBH & $3\,M_\odot < m_{2,\mathrm{src}} \leq m_{1,\mathrm{src}} < 100\,M_\odot$ &
            \result{\oThreeRealBBH} & \result{\oThreeSemiBBH} & \result{\oFourLowBBH} & \result{\oFourHighBBH} \\
         \hline
         \multirow{9}{*}{other} & \multirow{5}{*}{BBH}
           & $3\,M_\odot \leq m_{2,\mathrm{src}} \leq m_{1,\mathrm{src}} \leq 10\,M_\odot$ &
            \result{\oThreeRealBBHa} & \result{\oThreeSemiBBHa} & \result{\oFourLowBBHa} & \result{\oFourHighBBHa} \\
         & & ($3\,M_\odot \leq m_{2,\mathrm{src}} \leq m_{1,\mathrm{src}}$) AND ($10\,M_\odot \leq m_{1,\mathrm{src}} \leq 20\,M_\odot$) &
            \result{\oThreeRealBBHb} & \result{\oThreeSemiBBHb} & \result{\oFourLowBBHb} & \result{\oFourHighBBHb} \\
         & & ($3\,M_\odot \leq m_{2,\mathrm{src}} \leq m_{1,\mathrm{src}}$) AND ($20\,M_\odot \leq m_{1,\mathrm{src}} \leq 35\,M_\odot$) &
            \result{\oThreeRealBBHc} & \result{\oThreeSemiBBHc} & \result{\oFourLowBBHc} & \result{\oFourHighBBHc} \\
         & & ($3\,M_\odot \leq m_{2,\mathrm{src}} \leq m_{1,\mathrm{src}}$) AND ($35\,M_\odot \leq m_{1,\mathrm{src}} \leq 50\,M_\odot$) &
            \result{\oThreeRealBBHd} & \result{\oThreeSemiBBHd} & \result{\oFourLowBBHd} & \result{\oFourHighBBHd} \\
         & & ($3\,M_\odot \leq m_{2,\mathrm{src}} \leq m_{1,\mathrm{src}}$) AND ($50\,M_\odot \leq m_{1,\mathrm{src}} \leq 100\,M_\odot$) &
            \result{\oThreeRealBBHe} & \result{\oThreeSemiBBHe} & \result{\oFourLowBBHe} & \result{\oFourHighBBHe} \\
        \cline{2-7}
         & \multirow{4}{*}{mass ratio} & $0.00 \leq q = m_2/m_1 < 0.25$ &
            \result{\oThreeRealqa} & \result{\oThreeSemiqa} & \result{\oFourLowqa} & \result{\oFourHighqa} \\
          & & $0.25 \leq q = m_2/m_1 < 0.50$ &
            \result{\oThreeRealqb} & \result{\oThreeSemiqb} & \result{\oFourLowqb} & \result{\oFourHighqb} \\
          & & $0.50 \leq q = m_2/m_1 < 0.75$ &
            \result{\oThreeRealqc} & \result{\oThreeSemiqc} & \result{\oFourLowqc} & \result{\oFourHighqc} \\
          & & $0.75 \leq q = m_2/m_1 < 1.00$ &
            \result{\oThreeRealqd} & \result{\oThreeSemiqd} & \result{\oFourLowqd} & \result{\oFourHighqd} \\
        \hline
    \end{tabular}
    }
\end{table*}

\end{turnpage}

I note that this estimate does not account for uncertainty in the astrophysical population (nor does it use a model that was actually fit to the O3 data) and only uses an estimate of the possible sensitivity during O4.
As of the time of writing, the detectors have not reached the projected O4 sensitivities, and therefore the detection rate will be lower than I state here.


\section{Implications for computational cost of Monte Carlo sensitivity estimates}
\label{sec:hopeless cuts}

A key ingredient in the inference of astrophysical populations from the set of observed signals is the sensitivity of a search to a particular population, or the fraction of a population that would be detected by a search.
This, in turn, depends on the probability that a search would detect a signal with a particular set of parameters ($\theta$) averaged over the noise properties throughout the experiment
\begin{equation}\label{eq:average detection}
    \left< P(\mathrm{det}|\theta) \right>_{t,n} = \int dt\, p(t) \int \mathcal{D}n\, p(n|t) P(\mathrm{det}|d=n+h(\theta))
\end{equation}
where $P(\mathrm{det}|d)$ is a deterministic function of the data (i.e., whether a given piece of data would end up within the catalog).
One typically assumes the distribution of times when events occur $p(t)$ is uniform and models the distribution of detector noise $p(n|t)$ by either assuming stationary Gaussian noise or using the actual noise produced by real detectors.
Inference for population properties then depends on integrals over the population (described by hyperparameters $\Lambda$)
\begin{equation}
    P(\mathrm{det}|\Lambda) = \int d\theta\, p(\theta|\Lambda) \left< P(\mathrm{det}|\theta)\right>_{t,n}
\end{equation}
These high-dimensional integrals are often estimated through injection campaigns in which searches process simulated signals drawn from a known reference distribution and report which are detected.

Current detectors produce flux-limited surveys.
Therefore, the majority of sources will be at high redshifts.
This implies that the majority of the population will be too far away to detect ($P(\mathrm{det}|\Lambda) \ll 1$).
In order to reduce the computational burden placed on searches, then, the population is often artificially truncated at reasonable redshifts just beyond the detectors' observable horizon.
However, additional thresholds on $\rho_\mathrm{opt}$ are also imposed in order to only select injections that searches have some hope of detecting, even if that hope is very small.
These ``hopeless cuts'' can dramatically reduce the number of injections that searches must process and \emph{de facto} approximate $P(\mathrm{det}|\rho_\mathrm{opt}(\theta) < \rho_\mathrm{cut}) = 0$.
It is therefore of some interest to know how to choose $\rho_\mathrm{cut}$; it should be as large as possible (to minimize computational cost) without significantly affecting the estimate of $P(\mathrm{det}|\Lambda)$.


\subsection{Analytic Estimates}
\label{sec:analytic estimates}

If one assumes searches' ability to detect signals is driven primarily by the observed filter response ($\rho_\mathrm{obs}$) and approximates this with a simple threshold on $\rho_\mathrm{obs}$, then the extent that a hopeless cut affects Monte Carlo estimates of search sensitivity can be phrased in terms of a truncation error
\begin{multline}
    \Delta P(\mathrm{det}|\Lambda) \equiv \epsilon \\ = \int\limits_{0}^{\rho_\mathrm{cut}} d\rho_\mathrm{opt}\, p(\rho_\mathrm{opt}|\Lambda) P(\rho_\mathrm{obs} \geq \rho_\mathrm{thr}|\rho_\mathrm{opt}) \label{eq:this one}
\end{multline}
where
\begin{equation}
    p(\rho_\mathrm{opt}|\Lambda) = \int d\theta\, p(\theta|\Lambda) \delta(\rho_\mathrm{opt} - \rho_\mathrm{opt}(\theta))
\end{equation}

Now, there remains the question of what an appropriate choice for $\epsilon$ might be.
The systematic (truncation) error in $P(\mathrm{det}|\Lambda)$ should be small compared to the statistical uncertainty from the finite Monte Carlo sample size.
Following the convention of~\citet{Farr:2019}, this implies
\begin{equation}\label{eq:requirement}
    \epsilon \ll \sigma = \frac{P(\mathrm{det}|\Lambda)}{\sqrt{N_\mathrm{eff}}}
\end{equation}
which is nothing more than the definition of the effective number of samples ($N_\mathrm{eff}$) within the Monte Carlo sum.
Similarly,~\citet{Farr:2019} and~\citet{Essick:2022} suggest that stable hierarchical Bayesian inference is only possible if $N_\mathrm{eff} \gtrsim 4 N_\mathrm{cat}$, where $N_\mathrm{cat}$ is the number of events in the catalog.\footnote{There is disagreement about this in the literature. See, e.g.,~\citet{Talbot:2023} for an alternate perspective.}
In general, one may wish to include some additional factor of safety ($f_\mathrm{fdg}$) so that $N_\mathrm{eff} = f_\mathrm{fdg} N_\mathrm{cat}$.
Additionally, because $P(\rho_\mathrm{obs} \geq \rho_\mathrm{thr}|\rho_\mathrm{opt})$ monotonically increases with $\rho_\mathrm{opt}$,
\begin{align}
    \epsilon
        & \leq P(\rho_\mathrm{obs} \geq \rho_\mathrm{thr}|\rho_\mathrm{cut}) \int\limits_{0}^{\rho_\mathrm{cut}} d\rho_\mathrm{opt}\, p(\rho_\mathrm{opt}|\Lambda) \nonumber \\
        & \leq P(\rho_\mathrm{obs} \geq \rho_\mathrm{thr}|\rho_\mathrm{cut}) \label{eq:eps upper limit}
\end{align}
This means one can guarantee Eq.~\ref{eq:requirement} is satisfied if
\begin{equation}\label{eq:rule of thumb}
    P(\rho_\mathrm{obs} \geq \rho_\mathrm{thr}|\rho_\mathrm{cut}) \ll \frac{P(\mathrm{det}|\Lambda)}{\sqrt{f_\mathrm{fdg} N_\mathrm{cat}}}
\end{equation}
Let us use expectations for the LVK's fourth observing run (O4) as an example.
With the assumed astrophysical distribution introduced in Sec.~\ref{sec:case study}, I find $P(\mathrm{det}|\Lambda) \approx 0.05$.
A reasonable (but optimistic) estimate for the size of the catalog at the end of O4 is $N_\mathrm{cat} = 500$ (see Sec.~\ref{sec:projections for O4} for more precise estimates), and I take $f_\mathrm{fdg} = 10$ to again be conservative.
Putting this together yields $P(\rho_\mathrm{obs} \geq \rho_\mathrm{thr} | \rho_\mathrm{opt}=\rho_\mathrm{cut}) < 7\times10^{-4}$, which corresponds to $\rho_\mathrm{cut} \sim 5.5$ based on the observed distribution of $\rho_{\mathrm{net},\phi}$ with $\rho_\mathrm{thr}=9$ and three detectors (Fig.~\ref{fig:3ifo}).

Note, however, this requirement may be too strict.
One can make $P(\mathrm{det}|\Lambda)$ arbitrarily small by changing the population to add more signals at very high redshifts.
This, in turn, would require a much smaller $\rho_\mathrm{cut}$ from Eq.~\ref{eq:rule of thumb}.
However, this need not be the case as $P(\rho_\mathrm{obs} \geq \rho_\mathrm{thr}|\rho_\mathrm{opt})$ will also decrease dramatically for those high-redshift systems.
That is, $\epsilon$ may be much smaller than the upper limit in Eq.~\ref{eq:eps upper limit}.
As such, requirements for $\rho_\mathrm{cut}$ based on Eq.~\ref{eq:rule of thumb} are sufficient but not necessary to guarantee $\epsilon \ll \sigma$.
More precise requirements must rely on direct experimentation.
Sec.~\ref{sec:case study} does exactly that.


\subsection{Demonstration of Monte Carlo sensitivity estimates}
\label{sec:case study}

\begin{figure}
    \begin{center}
        \includegraphics[width=1.0\columnwidth]{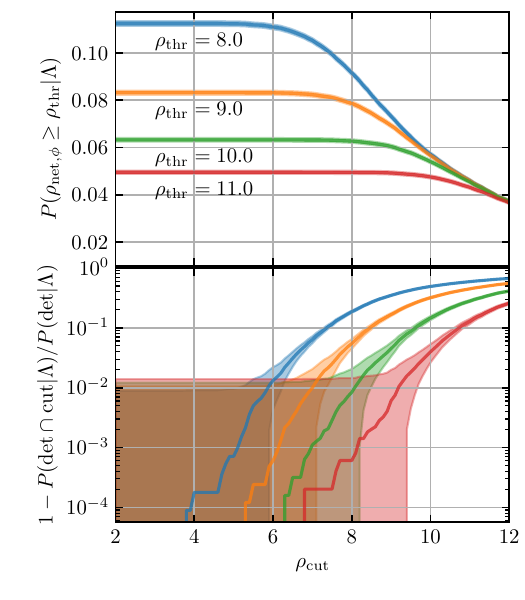}
    \end{center}
    \caption{
        Semianalytic approximations of the detection probability assuming a fixed population with different detection thresholds ($\rho_\mathrm{thr}$) and hopeless cuts ($\rho_\mathrm{cut}$) on $\rho_{\mathrm{net},\mathrm{opt}}$.
        (\emph{top}) the detection probability modeled as a cut on $\rho_{\mathrm{net},\phi}$ and (\emph{bottom}) the difference between the approximation to the detection probability with vs. without a hopeless cut.
        Shaded regions approximate $1\sigma$ uncertainty from the finite number of Monte Carlo samples.
        Calibration uncertainty is thought to limit the precision in the rate inferred for a fixed population to within $O(1\%)$ relative uncertainty.
    }
    \label{fig:monte carlo summary}
\end{figure}

\begin{figure*}
    \begin{center}
        \includegraphics[width=1.0\textwidth, clip=True, trim=0.0cm 1.3cm 0.0cm 0.0cm]{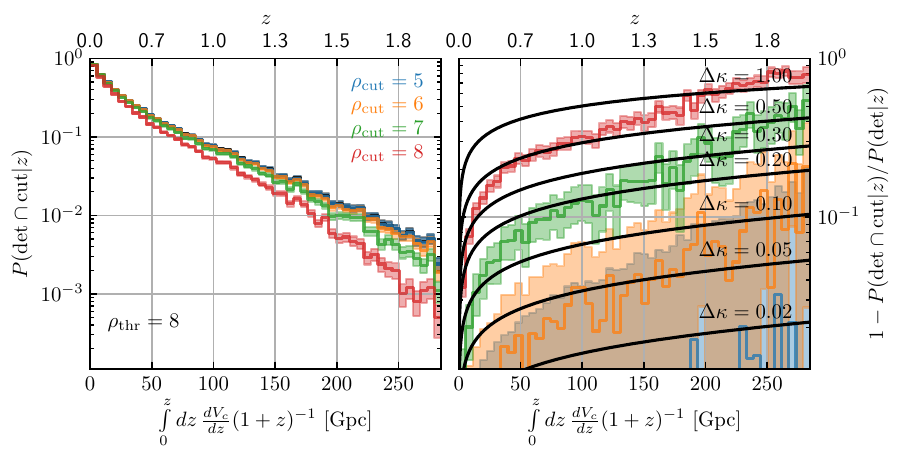}
        \includegraphics[width=1.0\textwidth, clip=True, trim=0.0cm 0.0cm 0.0cm 0.8cm]{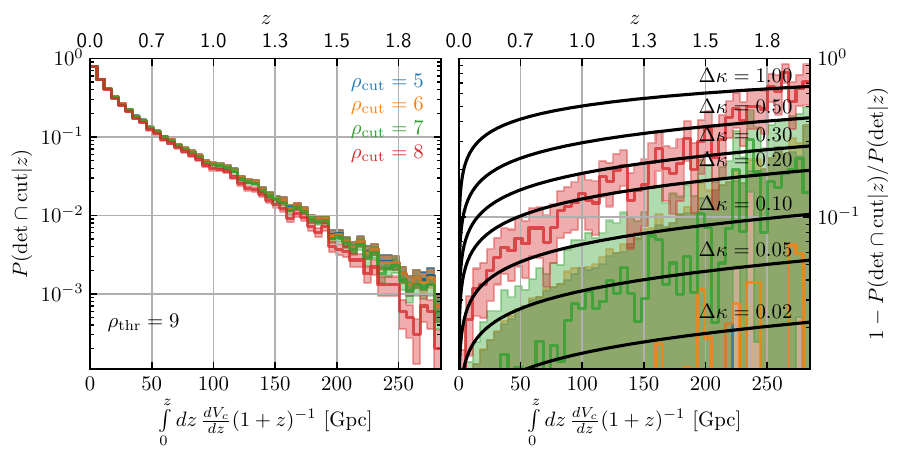}
    \end{center}
    \caption{
        (\emph{left}) Detection probability as a function of redshift for several detection thresholds (\emph{top}: $\rho_\mathrm{thr}=8$ and \emph{bottom}: $\rho_\mathrm{thr}=9$) and hopeless cuts (\emph{blue, orange, green, red}: $\rho_\mathrm{cut} = 5, 6, 7, 8$) along with (\emph{right}) ratios of the estimated detection probability with a hopeless cut to the detection probability without any hopeless cut.
        Separate Monte Carlo sets are drawn within each bin (uniformly spaced in comoving volume and source-frame time), and shaded regions approximate $1\sigma$ uncertainty from the finite Monte Carlo sample size.
        Black lines show the amount by which truncation error in estimates of $P(\mathrm{det}|z)$ could be compensated by redshift evolution of the form $dN/dz \sim (1+z)^{\kappa-1} (dV_c/dz)$.
    }
    \label{fig:monte carlo details}
\end{figure*}

As a concrete demonstration, I draw a large sample from a simple reference population and consider the effect of different hopeless cuts on the resulting sensitivity estimates.
This simulation used a three detector network (HLV) with nominal design sensitivities for the advanced detectors.\footnote{The projected advanced LIGO design sensitivity was taken from~\citet{aLIGO-design-psd}, and the advanced Virgo design sensitivity is the same one used in Fig. 4 of~\citet{Essick:2017}.}
The reference population is
\begin{itemize}
    \item a power-law in source-frame primary mass between 10--50 $M_\odot$ with an exponent of -1.3, 
    \item a power-law in source-frame secondary mass between 10 $M_\odot$ and the primary mass with an exponent of +4,
    \item zero spins, and 
    \item uniform in comoving volume and source-frame time assuming a Plank~\cite{Planck:2018} flat $\Lambda$CDM cosmology out to $z=2$. 
\end{itemize}
Fig.~\ref{fig:monte carlo summary} summarizes the result with a few detection thresholds based on $\rho_{\mathrm{net},\phi}$.
Fig.~\ref{fig:monte carlo details} breaks this down into a more detailed picture, focusing on the population property most likely to be affected by a poor choice of $\rho_\mathrm{cut}$: the merger rate's evolution with redshift.

With this population, \result{$P(\mathrm{det}|\Lambda) \sim O(0.05)$} and can be accurately estimated for a range of $\rho_\mathrm{cut}$.
In fact, Fig.~\ref{fig:monte carlo summary} suggests that we can estimate $P(\mathrm{det}|\Lambda)$ to better than $O(1\%)$ relative uncertainty with \result{$\rho_\mathrm{cut} \lesssim 7$} as long as \result{$\rho_\mathrm{thr} \gtrsim 9$}.
Note that this requirement is larger than what was obtained in Sec.~\ref{sec:analytic estimates}.
Again, that is expected because Sec.~\ref{sec:analytic estimates} derives a sufficient but not necessary requirement.

Now, bounding the error in $P(\mathrm{det}|\Lambda)$ for a single population is necessary to accurately estimate the astrophysical rate of mergers assuming that population.
One may wish to make the systematic uncertainty in the rate for this single population (and therefore the systematic error in $P(\mathrm{det}|\Lambda)$) smaller than the systematic uncertainty from detector calibration, which is thought to be a relative uncertainty of $O(1\%)$.
By this criterion, $\rho_\mathrm{cut} \sim 7$ may be acceptable for $\rho_\mathrm{thr} = 9$.
However, choosing $\rho_\mathrm{cut}$ based on this criterion may overlook the important features that become apparent when the shape of the distribution is inferred rather than assumed.
Indeed, large $\rho_\mathrm{cut}$ will preferentially impact injections with large redshifts and may introduce biases in the inference of the merger rate's redshift evolution.

Fig.~\ref{fig:monte carlo details} investigates this in more detail.
Specifically, it breaks down $P(\mathrm{det}|\Lambda)$ into subpopulations, each spanning a small range in redshift.
By comparing the impact of $\rho_\mathrm{cut}$ on these different subpopulations, we can evaluate the possibility that it will bias the inferred astrophysical population as a function of redshift.
Fig.~\ref{fig:monte carlo details} shows that the hopeless signal-to-noise ratio cuts preferentially affect the highest redshifts (systems with lowest $\rho_{\mathrm{net},\mathrm{opt}}$).
While $P(\mathrm{det}|z)$ always decreases with $z$, it decreases faster with larger $\rho_\mathrm{cut}$.

Note that this truncation error maps directly onto the inferred redshift evolution.
That is, astrophysical inference depends primarily on the expected number of events through an integral of the form
\begin{equation}
    \mathrm{E}[N_\mathrm{cat}|\Lambda] \propto \int dz\, P(\mathrm{det}|z) p(z|\Lambda)
\end{equation}
and many redshift-evolution models (e.g.,~\citet{Fishbach:2018, Fishbach:2021, GWTC-2, GWTC-3}, and others) adopt the form
\begin{equation}\label{eq:redshift distribution}
    p(z|\Lambda) \propto (1+z)^{\kappa-1} \frac{dV_c}{dz}
\end{equation}
so that
\begin{equation}\label{eq:expected number}
    \mathrm{E}[N_\mathrm{cat}|\Lambda] \propto \int \left(dz\, \frac{dV_c}{dz} (1+z)^{-1}\right) P(\mathrm{det}|z) (1+z)^\kappa
\end{equation}
A systematic underestimation of $P(\mathrm{det}|z)$ can be compensated by an overestimation of $\kappa$ as long as the integral remains roughly constant.
Fig.~\ref{fig:monte carlo details} shows exactly this with bins spaced uniformly in comoving volume and source-frame time (the measure of Eq.~\ref{eq:expected number}).
More flexible models of redshift evolution (e.g.,~\citet{Callister:2023} and~\citet{Edelman:2022}) may even trace out the specific bias in $P(\mathrm{det}|z)$ from truncation error induced by $\rho_\mathrm{cut}$.

Fig.~\ref{fig:monte carlo details} shows that we might expect biases of \result{$\Delta \kappa \lesssim 0.2$ with $\rho_\mathrm{thr}=8$ and $\rho_\mathrm{cut}=6$}.
However, if \result{$\rho_\mathrm{thr}=9$, then we expect comparable biases with $\rho_\mathrm{cut}=7$}.

Now, GWTC-3~\cite{GWTC-3} reports $\kappa = 2.9^{+1.7}_{-1.8}$.
If we expect a roughly 5-fold increase in the catalog size by the end of O4 ($N_\mathrm{cat} \sim 500$, see also Sec.~\ref{sec:projections for O4}), then we might expect statistical uncertainties $\Delta \kappa = \pm 0.78$ if the constraint scales as $N_\mathrm{cat}^{-1/2}$ or $\Delta \kappa = \pm 0.35$ if it scales as $N_\mathrm{cat}^{-1}$.
It is known that constraints on $\kappa$ scale faster than $N_\mathrm{cat}^{-1/2}$ because the additional observations come from larger redshifts (assuming increased detector sensitivity), but it is unlikely to scale faster than $N_\mathrm{cat}^{-1}$.
Therefore, this suggests that we must have \result{$\rho_\mathrm{cut} < 7$ if $\rho_\mathrm{thr}=9$ to keep possible biases in $\kappa$ smaller than the best-case statistical uncertainty expected at the end of O4}.
Note that this comparable to the requirement to match calibration uncertainty in the rate for a fixed population and is looser than the estimate from Sec.~\ref{sec:analytic estimates}.


\subsection{Impact of PSD variability}
\label{sec:psd variability}

The preceding estimates suggest that $\rho_\mathrm{cut} \sim 7$ may be marginally tolerable given $\rho_\mathrm{thr} = 9$ and that $\rho_\mathrm{cut} \sim 6$ should be more than sufficient for our current needs (resolve rates better than calibration uncertainty and limit bias in the inferred redshift evolution).
However, all the preceding estimates for appropriate $\rho_\mathrm{cut}$ were based on the assumption that the PSD is stationary.
If the PSD varies relative to the reference used to define an injection set, then the $\rho_\mathrm{opt}$ seen by searches will similarly vary relative to the estimated $\rho_\mathrm{opt}^{(\mathrm{ref})}$ used when performing the hopeless cut.
One should therefore build in additional factors of safety in $\rho_\mathrm{cut}^{(\mathrm{ref})}$ to account for expected PSD variability if they wish to avoid truncation errors during any part of the run.

It is generally risky to predict the variability of the PSD before the beginning of a run, but previous runs have seen the binary neutron star (BNS) range vary by $\sim 10$--$20\%$ (see, e.g., Fig. 3 of~\citet{Buikema:2020}).
Using this as a baseline, an injection estimated to have $\rho_\mathrm{opt}^{(\mathrm{ref})} = 5$ with a reference PSD may actually have $\rho_\mathrm{opt} \gtrsim 6$ with respect to the actual PSD at the time of the injection.

As such, we conclude that $\rho_\mathrm{cut}^{(\mathrm{ref})} = 6$ (as estimated with a reference PSD) is marginally sufficient ($\rho_\mathrm{opt}$ may actually be $\gtrsim 7$ given PSD variability).
This matches the empirical observation that the inferred redshift evolution from GWTC-3 does not change if one lowers $\rho_\mathrm{cut}^{(\mathrm{ref})}$ below $\sim 6$ (estimated with a reference PSD)~\cite{Callister:2023-dcc, Golomb:2023-dcc}.
The bias in redshift evolution observed in these studies at $\rho_\mathrm{cut}^{(\mathrm{ref})} = 7$ can be $\Delta \kappa \gtrsim 1$, which quantitatively agrees with our estimates from Sec.~\ref{sec:case study} assuming $\rho_\mathrm{thr}=9$ and $\rho_\mathrm{cut}=8$.

Taken together with an abundance of caution, \result{setting $\rho_\mathrm{cut}^{(\mathrm{ref})} \sim 5$ for a detection threshold $\rho_\mathrm{thr}=9$ based on $\rho_{\mathrm{net},\phi}$ should be sufficient for O4}.
The expectation is that $\rho_\mathrm{opt}$ at the time of the injection will not be scattered above $\sim 6$ due to PSD variability.
Larger $\rho_\mathrm{cut}^{(\mathrm{ref})}$ could be acceptable if the PSD was known more precisely or if one was willing to risk biases in the inferred redshift evolution more than $O(10\%)$ of the expected statistical uncertainty after O4.


\section{Discussion}
\label{sec:conclusion}

I have presented an overview of semianalytic sensitivity estimates for catalogs of coalescing binaries observed through GWs.

By first deriving the expected distribution of the response of matched filters in stationary Gaussian noise, including correlations between neighboring templates, I showed that \result{one can accurately approximate the survival function of observed filter responses maximized across a template bank with a simple model for the response of the best-matching template}.
This approximation holds more precisely for intrinsically loud signals and high detection thresholds, but it is already nearly exact for realistic detection thresholds corresponding to real searches ($\rho_\mathrm{thr} \gtrsim 9$).

I then showed that \result{constructing semianalytic sensitivity estimates based on the observed signal-to-noise ratio ($\rho_{\mathrm{net},\phi}$) results in a better fit to the distribution of real detected injections from O3 compared to models based on selection with the optimal signal-to-noise ratio ($\rho_{\mathrm{net},\mathrm{opt}}$}).
This is because $\rho_{\mathrm{net},\phi}$ contains additional variance by incorporating the effects of detector noise, and this is important when matching the true selection performed by searches.
While the exact behavior differs from search to search, there are consistent trends in the approximate detection threshold as a function of binary masses, with higher masses requiring larger detection thresholds.
Regardless, \result{approximating the selection with a threshold $\rho_{\mathrm{net},\phi} \geq 10$ produces a reasonable approximation to the behavior of real searches across the entire mass range.}
Finally, I employ this knowledge to compare distributions of detected signals in O3 with projections for O4, finding comparable ratios of different types of systems in the detected populations from each run.

After demonstrating that semianalytic estimates are reliable, I turned to the question of how to generate them most efficiently by selecting a hopeless signal-to-noise ratio cut.
After accounting for the possibility of PSD variability, \result{I recommend a hopeless cut $\rho_\mathrm{cut}=5$ on the optimal signal-to-noise ratio assuming a detection threshold on the observed signal-to-noise ratio of $\sim 9$}.

The semianalytic approximation is not perfect (Fig.~\ref{fig:detected corner} and Table~\ref{tab:o4 projections} show some disagreement in the detected mass and spin distributions when a single threshold is applied for a relatively broad mass range), but it nevertheless provides a powerful way to capture the relevant physics of GW detector sensitivity in terms of a single number that can generalize to arbitrary detector networks and/or PSDs.
Future work may focus on improving the semianalytic approximation by explicitly incorporating the dependence of $\rho_\mathrm{thr}$ on single-event parameters (like $\mathcal{M}_\mathrm{det}$; Fig.~\ref{fig:threshold by mass}) and/or develop analytic models for the distributions of goodness-of-fit tests employed within searches (see, e.g.,~\citet{Tsukada:2023}).
This may provide a physics-driven approach to select which additional features describing individual events must be included to develop a sufficient set of statistics to fully capture the sensitivity of matched-filter searches.
Additionally, a more detailed investigation into the impact of PSD variability may also be warranted.
Although this does not seem to significantly impact the emulators studied in Sec.~\ref{sec:real injections}, which use fixed reference PSDs, one may wish to measure PSDs throughout a run and construct a mixture model of emulators~\cite{Essick:2021}, each using separate measured PSDs for short sections of the run.


\acknowledgments

I am very grateful to Maya Fishbach, Tom Callister, Jacob Golomb, Will Farr, Daniel Holz, Divya Singh, and Surabhi Sachdev for useful conversations during the preparation of this manuscript.
I am also grateful for computational resources provided by the LIGO Laboratory and supported by National Science Foundation Grants PHY-0757058 and PHY-0823459.
This material is based upon work supported by NSF's LIGO Laboratory which is a major facility fully funded by the National Science Foundation.

I am supported by the Natural Sciences \& Engineering Research Council of Canada (NSERC).

This work would not have been possible without \texttt{numpy}~\cite{numpy}, \texttt{scipy}~\cite{scipy}, and \texttt{matplotlib}~\cite{matplotlib}.
Monte Carlo samples were drawn with \texttt{gw-detectors}~\cite{gw-detectors}, \texttt{gw-distributions}~\cite{gw-distributions}, and \texttt{monte-carlo-vt}~\cite{monte-carlo-vt}.


\bibliography{refs}

\end{document}